\documentclass [preprint,showpacs,nofootinbib,superscriptaddress]{revtex4}
\usepackage{bbm}
\usepackage{verbatim}
\usepackage{slashed}
\usepackage{epsfig}
\usepackage{graphicx}
\usepackage{amsmath}
\usepackage{mathrsfs}
\usepackage{bm}

\newcommand{\re}{\mathop{\mathrm{Re}}\nolimits}

\begin{document}

\preprint{DESY~15-092\hspace{12cm}ISSN~0418-9833}
\preprint{June 2015\hspace{15.2cm}}
\boldmath
\title{Relativistic corrections to $J/\psi$ polarization in photo- and
hadroproduction}
\unboldmath


\author{Zhi-Guo He}
\affiliation{
{\normalsize II. Institut f\"ur Theoretische Physik, Universit\"at Hamburg,}\\
{\normalsize Luruper Chaussee 149, 22761 Hamburg, Germany}
}
\author{Bernd A. Kniehl}
\affiliation{
{\normalsize II. Institut f\"ur Theoretische Physik, Universit\"at Hamburg,}\\
{\normalsize Luruper Chaussee 149, 22761 Hamburg, Germany}
}

\date{\today}

\begin{abstract}
We systematically calculate the relativistic corrections to the polarization
variables of prompt $J/\psi$ photoproduction and hadroproduction using the
factorization formalism of nonrelativistic QCD.
Specifically, we include the ${}^3\!S_1^{[1]}$ and ${}^3\!P_{J}^{[1]}$
color-singlet and the ${}^3\!S_1^{[8]}$, ${}^1\!S_{0}^{[8]}$, and
${}^3\!P_{J}^{[8]}$ color-octet channels as well as the effects due to the
mixing between the ${}^3\!S_1^{[8]}$ and ${}^3\!D_1^{[8]}$ channels.
We provide all the squared hard-scattering amplitudes in analytic form.
Assuming the nonrelativistic-QCD long-distance matrix elements to satisfy the
velocity scaling rules, we find the relativistic corrections to be appreciable,
especially at small transverse momentum $p_T$ and large inelasticity $z$.
The results obtained here and in our previous work on the unpolarized yield
[Phys.\ Rev.\ D {\bf 90}, 014045 (2014)] will help to render global analyses of
prompt $J/\psi$ production data more complete and hopefully to shed light on
the $J/\psi$ polarization puzzle.
\end{abstract}


\maketitle
\section{Introduction}

Through concerted efforts from both the experimental and theoretical sides,
we have deepened our understanding of the mechanism of $J/\psi$ production
at high energies in recent years (for a review, see
Ref.~\cite{Brambilla:2010cs} and references cited therein).
Yet, there are still major challenges to the theoretical models.
One of them is the long-standing $J/\psi$ polarization puzzle.
The polarization of promptly produced $J/\psi$ mesons, which are produced
either directly or via the feed down from higher excited states such as
$\chi_{cJ}$ ($J=0,1,2$) and $\psi^\prime$ mesons, was measured by several
experiments in different environments, including CDF at the Fermilab
Tevtron \cite{Affolder:2000nn,Abulencia:2007us}; HERA-B \cite{Abt:2009nu}, 
ZEUS \cite{Chekanov:2009ad}, and H1 \cite{Aaron:2010gz} at DESY HERA; PHENIX at
BNL RHIC \cite{Adare:2009js}; and
ALICE \cite{Abelev:2011md}, CMS \cite{Chatrchyan:2013cla}, and LHCb
\cite{Aaij:2013nlm} at CERN LHC.
Unfortunately, these measurements could not yet be explained by theoretical
analyses in a way consistent with the world data on the unpolarized $J/\psi$
yield.
The production of heavy-quarkonium states, like the $J/\psi$ meson, involves
the creation of the heavy-quark pair ($Q\bar{Q}$) and its subsequent
transition to the hadronic bound state. 
Different ways of treating these two parts result in different models.
Polarization variables are more sensitive to the fine details of the
hadronization of the $Q\bar{Q}$ pair than production rates and, therefore,
play a key role in testing the theoretical models.

Due to the fact that the heavy-quark mass $m_Q$ is much larger than the
asymptotic scale parameter of QCD $\Lambda_\mathrm{QCD}$, the heavy-quarkonium
state can be approximately treated as a nonrelativistic system.
The factorization formalism based on the effective field theory of
nonrelativistic QCD (NRQCD) \cite{Caswell:1985ui,Lepage:1992tx,Bodwin:1994jh}
is nowadays considered to be the most favorable theoretical approach to study
heavy-quarkonium production and decay.
In this framework, the heavy-quarkonium production cross sections are
factorized into process-dependent short-distance coefficients (SDCs) and
supposedly universal long-distance matrix elements (LDMEs)
\cite{Bodwin:1994jh}.
The SDCs, which describe the production of the $Q\bar{Q}$ pair at energy scales
$2m_{Q}$ or larger, can be calculated perturbatively through expansions in the
strong-coupling constant $\alpha_s$.
The LDMEs, which measure the probability of the hadronization of the
$Q\bar{Q}$ pair, are weighted by definite powers of the relative velocity $v$
of the heavy quarks in the heavy-meson rest frame \cite{Lepage:1992tx}.
In this way, theoretical calculations are organized as double expansions in
$\alpha_s$ and $v$.
Quantum corrections come as terms of higher orders in $\alpha_s$ and
relativistic corrections as terms of higher orders in $v^2\approx\alpha_s$.
Thus, next-to-leading-order (NLO) results include corrections of orders
$\mathcal{O}(\alpha_s)$ and $\mathcal{O}(v^2)$ relative to the leading-order
(LO) result.
Next-to-next-to-leading-order results also include terms of relative orders
$\mathcal{O}(\alpha_s^2)$, $\mathcal{O}(\alpha_sv^2)$ , $\mathcal{O}(v^4)$,
and so on.
A crucial difference between the NRQCD factorization formalism and the
conventional color-singlet (CS) model is that the former allows for the
$Q\bar{Q}$ pair to be in any possible Fock state $n={}^{2S+1}\!L_{J}^{[a]}$
with total spin $S$, orbital angular momentum $L$, total angular momentum $J$,
and color multiplicity $a=1,8$, while the latter is restricted to the CS state,
with $a=1$, sharing the quantum numbers $S$, $L$, and $J$ with the heavy meson.
The color-octet (CO) mechanism, {\it i.e.}, the appearance of CO states, with
$a=8$, is a distinctive feature of NRQCD factorization.

Two decades after the introduction of NRQCD factorization \cite{Bodwin:1994jh},
all the relevant observables of prompt $J/\psi$ production are available at
NLO in $\alpha_s$.
In particular, these include yield \cite{Butenschoen:2009zy} and polarization
\cite{Butenschoen:2011ks} in photoproduction, yield
\cite{Ma:2010vd,Ma:2010yw,Butenschoen:2010rq} and polarization
\cite{Butenschoen:2012px,Chao:2012iv,Gong:2012ug,Shao:2014fca,Shao:2014yta} in
hadroproduction, and observables in other production modes
\cite{Klasen:2004tz,Klasen:2004az,Butenschoen:2011yh}. 
Since the LDMEs of the $\eta_c$ meson are related to those of the $J/\psi$
meson via heavy-quark spin symmetry, the $\eta_c$ yield
\cite{Butenschoen:2014dra,Han:2014jya,Zhang:2014ybe} must be included in this
list, too.
The $J/\psi$ LDMEs determined at NLO in NRQCD via a global fit to
measurements of the unpolarized $J/\psi$ yields in photoproduction,
hadroproduction, and other production modes \cite{Butenschoen:2011yh} lead to
predictions of $J/\psi$ polarization in hadroproduction
\cite{Butenschoen:2012px} that are incompatible with measurements at the
Tevatron \cite{Affolder:2000nn,Abulencia:2007us} and the LHC
\cite{Chatrchyan:2013cla,Aaij:2013nlm}.
On the other hand, $J/\psi$ LDMEs determined only from hadroproduction data
\cite{Chao:2012iv,Gong:2012ug} are incompatible with data from photoproduction
and other production modes \cite{Butenschoen:2012qr}.
In other words, the long-standing $J/\psi$ polarization crisis of NRQCD, which
had already been observed at LO \cite{Braaten:1999qk}, is substantiated at
NLO in $\alpha_s$.
This crisis has recently been aggravated by the observation
\cite{Butenschoen:2014dra} that all the up-to-date $J/\psi$ LDME sets
\cite{Butenschoen:2011yh,Chao:2012iv,Gong:2012ug,Bodwin:2014gia} lead to NLO
NRQCD predictions for the $\eta_c$ yield in hadroproduction that significantly
overshoot the recent LHCb measurement \cite{Aaij:2014bga}.
Unfortunately, attempts \cite{Han:2014jya} to reconcile the measured $\eta_c$
and $J/\psi$ yields of hadroproduction by resorting to pre-LHC LDMEs
\cite{Ma:2010yw} fail to describe the $J/\psi$ polarization measured at the
Tevatron \cite{Shao:2014yta} and the LHC \cite{Han:2014jya} in the regime of
large $p_T$ values and central rapidities $y$.
The analysis of Ref.~\cite{Zhang:2014ybe} also misses its goal of achieving a
coherent description of the $J/\psi$ and $\eta_c$ yields and the $J/\psi$
polarization in prompt hadroproduction.
Obviously, NRQCD factorization presently faces severe challenges with regard to
the predicted universality of the LDMEs at NLO in $\alpha_s$.

To resolve the $J/\psi$ polarization puzzle and to clarify the universality
problem of the NRQCD LDMEs, we systematically calculated, in our previous work
\cite{He:2014sga}, the relativistic corrections, of relative order
$\mathcal{O}(v^2)$, to the unpolarized $J/\psi$ yields in photoproduction and
hadroproduction by including both the direct and feed-down contributions.
We found that the $\mathcal{O}(v^2)$ corrections are appreciable, except for
the CS ${}^3\!S_1^{[1]}$ channel of hadroproduction, and that their line shapes
greatly differ between photoproduction and hadroproduction, which may offer a
chance to improve the goodness of the state-of-the-art determinations of the
LDMEs.
In this paper, we take the next logical step by calculating the
$\mathcal{O}(v^2)$ corrections to the $J/\psi$ polarization observables in
photoproduction and hadroproduction.
All these $\mathcal{O}(v^2)$ corrections must be included on top of the
respective $\mathcal{O}(\alpha_s)$ corrections
\cite{Butenschoen:2009zy,Butenschoen:2011ks,Ma:2010vd,Ma:2010yw,%
Butenschoen:2010rq,Butenschoen:2012px,Chao:2012iv,Gong:2012ug,Shao:2014fca,%
Shao:2014yta}
to render the NLO predictions complete.

We organize the remainder of this paper as follows.
In Sec.~\ref{sec:two}, we shall briefly describe how to calculate the SDCs for
the polarization parameters in the direct and feed-down processes of
photoproduction and hadroproduction.
In Sec.~\ref{sec:three}, we shall present and discuss our numerical results.
A brief summary will be given in Sec.~\ref{sec:four}.
All the relevant analytic expressions will be collected in
the Appendix.

\section{NRQCD factorization formula}
\label{sec:two}

The polarization of the $J/\psi$ meson is measured through the angular
distribution of its leptonic decay, $J/\psi\to l^++l^-$, which may be
parametrized as
\begin{equation}
W(\theta,\phi)=1+\lambda_\theta\cos^2\theta+\lambda_\phi\sin^2\theta\cos(2\phi)+
\lambda_{\theta\phi}\sin(2\theta)\cos\phi,
\label{eq:w}
\end{equation}
where $\theta$ and $\phi$ are the polar and azimuthal angles of lepton $l^+$ in
the $J/\psi$ rest frame, respectively, which depend on the choice of coordinate
frame.
For example, in the recoil or $s$-channel helicity frame, the $z$ axis is
chosen to be the $J/\psi$ flight direction.
The parameters $\lambda_\theta$, $\lambda_\phi$, and $\lambda_{\theta\phi}$ in
Eq.~(\ref{eq:w}) may be related to the helicity density matrix
$d\sigma_{\lambda\lambda^\prime}$, which describes the interferences between the
states of helicity $\lambda$ and $\lambda^\prime$ in $J/\psi$ production
\cite{Beneke:1998re}
\begin{equation}
\lambda_\theta=\frac{d\sigma_{1,1}-d\sigma_{0,0}}{d\sigma_{1,1}+d\sigma_{0,0}},
\quad
\lambda_\phi=\frac{d\sigma_{1,-1}}{d\sigma_{1,1}+d\sigma_{0,0}},\quad
\lambda_{\theta\phi}=\frac{\sqrt{2}\re d\sigma_{1,0}}{d\sigma_{1,1}+d\sigma_{0,0}}.
\label{eq:lam}
\end{equation}

Invoking the Weizs\"{a}cker--Williams approximation and the factorization
theorems of the QCD parton model and NRQCD \cite{Bodwin:1994jh}, we may write
the hadronic helicity density matrices of both the direct and feed-down
processes in the general form
\begin{eqnarray}\label{xs}
\lefteqn{d\sigma_{\lambda\lambda^{\prime}}(A+B\to J/\psi+X)
=\sum_{i,j,H} \int d x_1 d y_1 d x_2\, f_{i/A}(x_1)f_{k/i}(y_1)f_{j/B}(x_2)}
\nonumber\\
&&{}\times d\hat{ \sigma}_{\lambda\lambda^{\prime}}(k+j\to H +X)
\mathrm{Br}(H\to J/\psi+X),
\end{eqnarray}
where $f_{i/A}(x)$ is the parton density function (PDF) of the parton $i$
in the hadron $A=p,\bar{p}$ or the flux function of the photon $i=\gamma$ in
the charged lepton $A=e^-,e^+$, $f_{j/i}(y_1)$ is $\delta_{ij}\delta(1-y_1)$ or
the PDF of the parton $j$ in the resolved photon $i=\gamma$,
$\mathrm{Br}(H\to J/\psi+X)$ is the branching ratio for
$H=J/\psi,\chi_{cJ},\psi^\prime$ with the understanding that
$\mathrm{Br}(J/\psi\to J/\psi+X)=1$, and
$d\hat{\sigma}_{\lambda\lambda^\prime}(i+j\to H +X)$ is the partonic helicity
density matrix.
Through relative order $\mathcal{O}(v^2)$, the latter may be
factorized as:\footnote{%
The spin-flip interactions are $\mathcal{O}(v^3)$ suppressed
\cite{Bodwin:2005gg}, so that NRQCD factorization still holds for the
helicity density matrix through relative order $\mathcal{O}(v^2)$.}
\begin{eqnarray}
d\hat{\sigma}_{\lambda\lambda^\prime}(i+j\to H +X)=\sum_{n}
\left(\frac{d F^{ij}_{\lambda\lambda^{\prime}}(n)}{m_c^{d_{\mathcal{O}(n)}-4}}
\langle\mathcal{O}^{H}(n)\rangle
+\frac{d G^{ij}_{\lambda\lambda^\prime}(n)}{m_c^{d_{\mathcal{P}(n)}-4}}
\langle\mathcal{P}^{H}(n)\rangle\right),
\end{eqnarray}
where $\mathcal{O}^{H}(n)$ is the four-quark operator pertaining to the
transition $n\to H$ at LO, with dimension $d_{\mathcal{O}(n)}$,
$\mathcal{P}^{H}(n)$ is related to its $\mathcal{O}(v^2)$ correction and
carries dimension $d_{\mathcal{P}(n)}=d_{\mathcal{O}(n)}+2$, and
$F^{ij}_{\lambda\lambda^\prime}(n)$ and $G^{ij}_{\lambda\lambda^\prime}(n)$ are
the appropriate SDCs of the partonic subprocess $i+j\to c\bar{c}(n)+X$.
The definitions of the $\mathcal{O}$ and $\mathcal{P}$ operators and some of
their properties may be found in Refs.~\cite{Bodwin:1994jh,He:2014sga}.
Working in the fixed-flavor-number scheme, the parton $i$ runs over the gluon
$g$ and the light quarks $q=u,d,s$ and antiquarks $\overline q$.

As in the unpolarized case \cite{He:2014sga}, the relevant partonic
subprocesses include
\begin{eqnarray}
g+\gamma &\to& c\bar{c}({}^3\!S_1^{[1,8]},{}^1\!S_0^{[8]},{}^3\!P_J^{[8]})+g,
\nonumber\\
q(\bar{q})+\gamma &\to& c\bar{c}({}^3\!S_1^{[8]},{}^1\!S_0^{[8]},{}^3\!P_J^{[8]})+q(\bar{q}),
\nonumber\\
g+g &\to& c\bar{c}({}^3\!S_1^{[1,8]},{}^1\!S_0^{[8]},{}^3\!P_J^{[1,8]})+g,
\nonumber\\
q(\bar{q})+g &\to& c\bar{c}({}^3\!S_1^{[8]},{}^1\!S_0^{[8]},{}^3\!P_J^{[1,8]})+q(\bar{q}),
\nonumber\\
\bar{q}+q &\to& c\bar{c}({}^3\!S_1^{[8]},{}^1\!S_0^{[8]},{}^3\!P_J^{[1,8]})+g.
\label{eq:sub}
\end{eqnarray}

The SDCs $F^{ij}_{\lambda\lambda^\prime}(n)$ and
$G^{ij}_{\lambda\lambda^\prime}(n)$ may still be calculated with the help of
the spinor projection method as explained in Ref.~\cite{He:2014sga} when the
polarization four-vector $\epsilon^\mu(\lambda)$ of the $c\overline{c}$ pair is
kept generic.
In the following, we only explain the differences in computing the helicity
density matrix elements with respect to Ref.~\cite{He:2014sga}.
We only need to discuss the nontrivial cases of $S=1$.  
In the case of direct $J/\psi$ production, the total spin $S$ of the $c\bar{c}$
pair can safely be identified with the spin $S$ of the $J/\psi$ meson through
$\mathcal{O}(v^2)$ because spin-flip effects occur only at $\mathcal{O}(v^3)$
\cite{Bodwin:2005gg}, and the index $L_z$ related to the orbital angular
momentum of the $c\bar{c}$ pair may be summed over.
Starting from the partonic scattering amplitude
$M_\lambda(n)=M(i+j\to c\bar{c}(n,\lambda)+X)$, where $\lambda$ is the helicity
related to the total spin $S$ of the $c\overline{c}$ pair, the helcity density
matrix element is thus evaluated as $\rho^{ij}_{\lambda\lambda^\prime}(n)
=\overline{\sum}M_\lambda(n)M^*_{\lambda^\prime}(n)$
by summing over the orbital angular momentum $L_z$ of the $c\bar{c}$ pair and
the spins and colors of the other outgoing partons, contained in system $X$,
and averaging over the spins and colors of the incoming partons $i$ and $j$.
Through $\mathcal{O}(v^2)$, the general Lorentz structure of
$\rho^{ij}_{\lambda\lambda^\prime}(n)$ implies the decomposition
\begin{equation}\label{rho}
\rho^{ij}_{\lambda\lambda^\prime}(n)=\sum_k
\left[A_k^{ij}(n)+\mathbf{q}^2B_k^{ij}(n)\right]
s_k^{\mu\nu}\epsilon^*_{\mu}(\lambda)\epsilon_{\nu}(\lambda^\prime),
\end{equation}
where $k$ labels the process-independent four-tensors
\begin{equation}
s_1^{\mu\nu}=g^{\mu\nu},\quad
s_2^{\mu\nu}=k_1^\mu k_1^\nu,\quad
s_3^{\mu\nu}=k_2^\mu k_2^\nu,\quad
s_4^{\mu\nu}=k_1^\mu k_2^\nu+k_2^\mu k_1^\nu,
\end{equation}
with $k_1$ and $k_2$ being the four-momenta of the incoming partons;
$A_k^{ij}(n)$ and $B_k^{ij}(n)$ are the SDCs, which are functions of the
partonic Mandelstam variables $s$, $t$, and $u$ and depend on the considered
partonic subprocess, but not on the choice of coordinate frame; and
$\mathbf{q}$ is the relative three-momentum within the $c\bar{c}$ pair.
Because $P\cdot\epsilon(\lambda)=0$, where $P$ is the total four-momentum of
the $c\bar{c}$ pair, and $P^2=4E_q^2$ \cite{He:2014sga},
$\epsilon^\mu(\lambda)$ also implicitly depends on $\mathbf{q}^2$. 
To obtain the complete results at $\mathcal{O}(v^2)$, the contractions between
$s_k^{\mu\nu}$ and $\epsilon^*_\mu(\lambda)\epsilon_\nu(\lambda^\prime)$ also
need to be expanded as series in $\mathbf{q}^2$ after choosing a coordinate
frame.
Writing
\begin{equation}
s_k^{\mu\nu}\epsilon^*_{\mu}(\lambda)\epsilon_{\nu}(\lambda^\prime)
=c^k_{\lambda\lambda^\prime}
+\mathbf{q}^2 d^k_{\lambda\lambda^\prime}+\mathcal{O}(\mathbf{q}^4),
\end{equation}
it is straightforward to obtain
\begin{eqnarray}
\frac{F^{ij}_{\lambda\lambda^\prime}(n)}{m_c^{d_{\mathcal{O}(n)}-4}}
&=&\frac{1}{2s}
\int d \mathrm{LIPS}\sum_k A_k^{ij}(n)c^k_{\lambda\lambda^\prime},
\nonumber\\
\frac{G^{ij}_{\lambda\lambda^\prime}(n)}{m_c^{d_{\mathcal{P}(n)}-4}}
&=&\frac{1}{2s}
\int d \mathrm{LIPS}\sum_k \left[A_k^{ij}(n)\left(K c^k_{\lambda\lambda^\prime}
+d^k_{\lambda\lambda^\prime}\right)+B_k^{ij}(n)c^k_{\lambda\lambda^\prime}
\right],
\end{eqnarray}
where the definition of the factor $K$ is the same as in
Ref.~\cite{He:2014sga}.
Note that $c^k_{\lambda\lambda^\prime}$ and $d^k_{\lambda\lambda^\prime}$ are
frame dependent.

The radiative decays $\chi_{cJ}\to J/\psi+\gamma$ are not only affected by the
E1 transition, but also by the higher-multipole M2 ($\chi_{c1}$ and
$\chi_{c2}$) and E3 ($\chi_{c2}$) ones.
A detailed study \cite{Faccioli:2011be} revealed that only the angular
distribution of the photon is very sensitive to the higher-multipole
contributions, while the latter are negligible for the angular distribution of
the subsequent $J/\psi\to l^++l^-$ decay.
Therefore, it is sufficient to calculate the helicity density matrix elements
of $J/\psi$ production from $\chi_{cJ}$ feed down by connecting the $J/\psi$
polarization four-vector to the total spin of the $c\bar{c}$ pair in the
helicity density matrix elements of $\chi_{cJ}$ production and multiplying the
outcome by the branching ratio of $\chi_{cJ}\to J/\psi+\gamma$.
At first sight, this is surprising because the Lorentz structure of the
helicity density matrix elements of $\chi_{cJ}$ production look more
complicated \cite{Kniehl:2000nn,Kniehl:2003pc}.\footnote{%
We caution the reader of the following misprints in Ref.~\cite{Kniehl:2000nn}:
In Eq.~(22), the terms proportional to
$\langle\mathcal{O}^{\chi_{c0}}({}^3\!S_1^{[8]})\rangle
B(\chi_{cJ}\to J/\psi+\gamma)$ with $J=1,2$ should be multiplied by the 
respective total-spin multiplicities $(2J+1)$.
Three lines below Eq.~(23), $|R_P^\prime(0)|$ should be squared.
The last factor in the first line of the expression for $c_2$ in Eq.~(A7)
should read $(7\hat{s}^2+\hat{s}\hat{t}+7\hat{t}^2)$.}
However, after summing over the polarization $J_z$ of the $\chi_{cJ}$ meson,
the helicity density matrix elements may indeed be expressed in the form of
Eq.~(\ref{rho}) as well.
To ensure that our simplified method is correct, we compared with previous LO
calculations \cite{Kniehl:2000nn,Shao:2012fs} to find agreement.  

We generate the Feynman diagrams by using the FeynArts package
\cite{Kublbeck:1990xc} and calculate the amplitude squares with the help of the
FeynCalc package \cite{Mertig:1990an}.
As for direct $J/\psi$ production, we reproduce the LO results for
$\rho^{ij}_{\lambda\lambda^\prime}(n)$ in Ref.~\cite{Beneke:1998re} by setting
$\mathbf{q}^2=0$, while our results for the $\mathcal{O}(v^2)$ corrections are
new.
As for $J/\psi$ production via feed down from $\chi_{cJ}$ mesons, the LO
helicity density matrix elements are found to agree with the results of
Ref.~\cite{Kniehl:2000nn} after simplifying the latter, but are represented
here for the first time in compact form, and the $\mathcal{O}(v^2)$ corrections
are again new. 
We recover our results in Ref.~\cite{He:2014sga} by summing over the helicity
indices $\lambda$ and $\lambda^\prime$.
In 
the Appendix,
we list all the new results for $A_k^{ij}(n)$ and
$B_k^{ij}(n)$ in analytic form.

\section{Phenomenological results}
\label{sec:three}

We are now in a position to investigate the phenomenological significance of
the $\mathcal{O}(v^2)$ corrections to the polarization parameters in prompt
$J/\psi$ photoproduction and hadroproduction.
In our numerical analysis, we use $m_c=1.5$~GeV, $\alpha=1/137.036$, the LO
formula for $\alpha_s^{(n_f)}(\mu_r)$ with $n_f=4$ active quark flavors and
asymptotic scale parameter $\Lambda_\mathrm{QCD}^{(4)}=215$~MeV
\cite{Pumplin:2002vw}, the CTEQ6L1 set for proton PDFs \cite{Pumplin:2002vw},
the photon flux function given in Eq.~(5) of Ref.~\cite{Kniehl:1996we} with
$Q_{\rm max}^2=2.5~\mathrm{GeV}^2$ \cite{Aaron:2010gz}, and the choice
$\mu_r=\mu_f=\sqrt{p_T^2+4m_c^2}$ for the renormalization and factorization
scales.
In contrast to the case of the unpolarized $J/\psi$ yield in
Ref.~\cite{He:2014sga}, the size of the $\mathcal{O}(v^2)$ corrections to the
$J/\psi$ polarization parameters for a given $c\bar{c}$ Fock state $n$ cannot
be directly estimated from the SDC ratios
$R_{\lambda\lambda^\prime}(n)=(dF_{\lambda\lambda^\prime}(n)/dx)/
(dG_{\lambda\lambda^\prime}(n)/dx)$, where $x$ is some variable, because the
$d\sigma_{\lambda\lambda^\prime}$ values enter as ratios in Eq.~(\ref{eq:lam}).
For instance, if we had $R_{00}(n)=R_{11}(n)$ for some $n$, then the
respective contribution to $\lambda_\theta$ would go unchanged upon the
inclusion of the $\mathcal{O}(v^2)$ corrections.
Therefore, we shall directly study the shifts induced in the polarization
parameters by including the $\mathcal{O}(v^2)$ corrections for each $n$
separately.
To standardize the numerical discussion, we quote the $\mathcal{O}(v^2)$
corrections due to the ${}^3\!S_1^{[8]}$--${}^3\!D_1^{[8]}$ mixing relative to
the corresponding LO results for $n={}^3\!S_{1}^{[8]}$.
We exclude from our considerations the trivial $J=0$ cases
$n={}^1\!S_0^{[8]}$ and $n={}^3\!P_{0}^{[1]}$.
In want of fitted values of the LDMEs $\langle\mathcal{P}^{H}(n)\rangle$, we
estimate them with the help of the velocity scaling rule 
$\langle\mathcal{P}^{H}(n)\rangle/\langle\mathcal{O}^{H}(n)\rangle=m_c^2v^2$
\cite{Lepage:1992tx} varying $v^2$ from 0 to 0.3 to account for uncertainties.
Furthermore, we limit ourselves to the direct production of $J/\psi$ mesons and
their production via the feed down from $\chi_{cJ}$ mesons.
In the latter case, we approximately take into account the kinematic effect on
the transverse momentum $p_T$ of the $J/\psi$ meson by setting
$p_T=p_T^{\chi_{cJ}}M_{J/\psi}/M_{\chi_{cJ}}$, with
$M_{J/\psi}=3.097$~GeV, $M_{\chi_{c1}}=3.511$~GeV, and
$M_{\chi_{c2}}=3.556$~GeV \cite{Agashe:2014kda}, which is justified since
$p_T\gg M_{\chi_{cJ}}-M_{J/\psi}$ in typical experimental situations.
In the case of $\gamma p$ collisions, we only consider direct photoproduction
for illustration because resolved photoproduction shares the partonic
subprocesses with hadroproduction, which is studied separately.

\begin{figure}
\begin{tabular}{cc}
\includegraphics[scale=0.80]{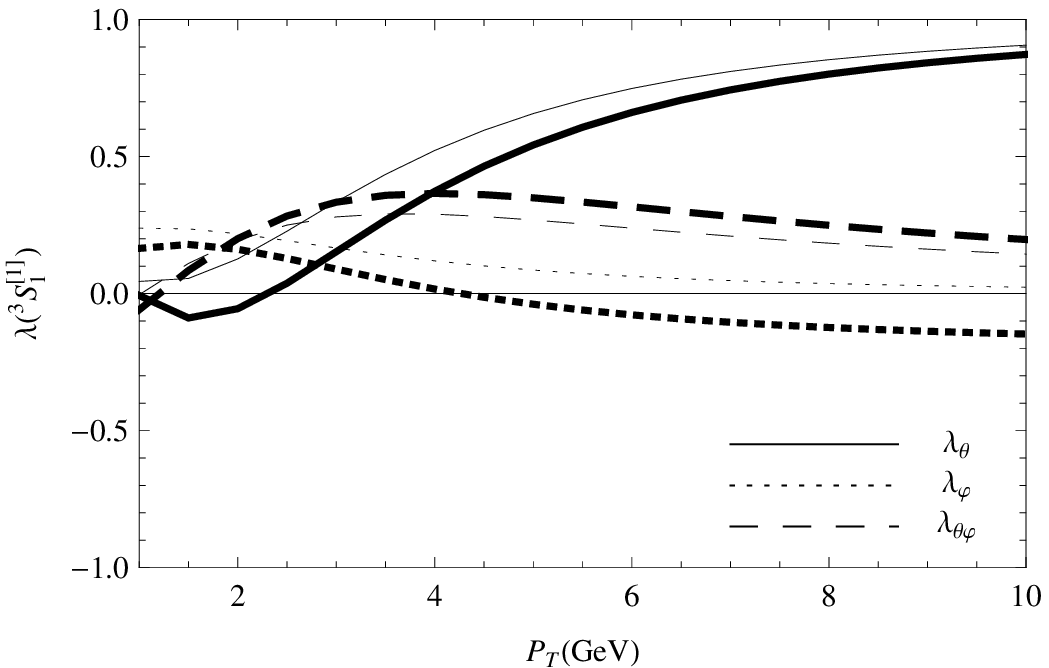}&
\includegraphics[scale=0.80]{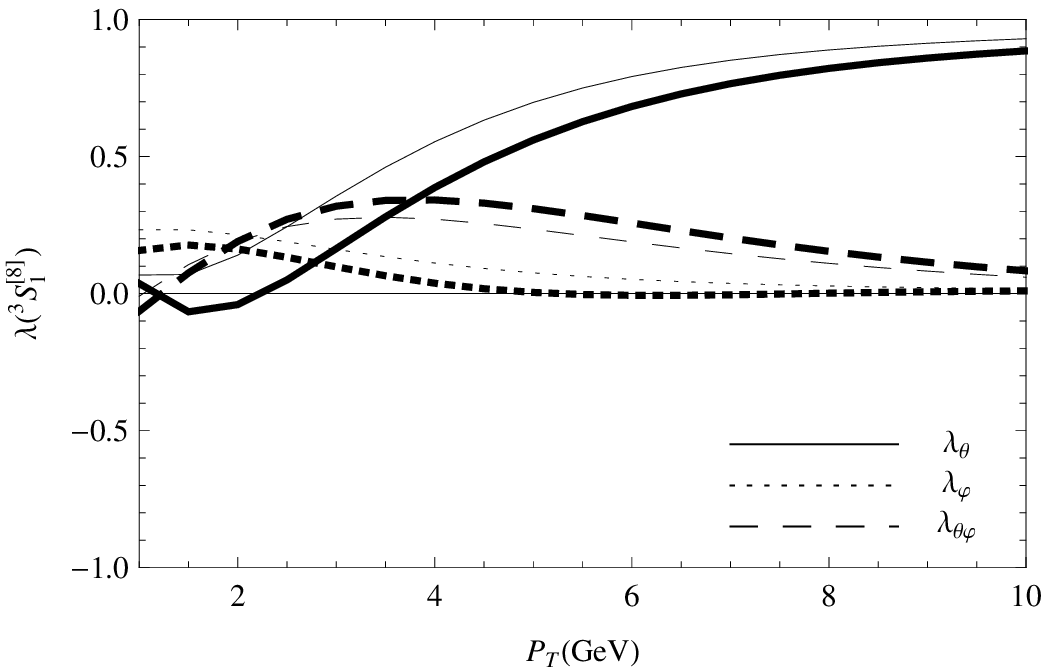}\\
(a)&(b)\\
\includegraphics[scale=0.80]{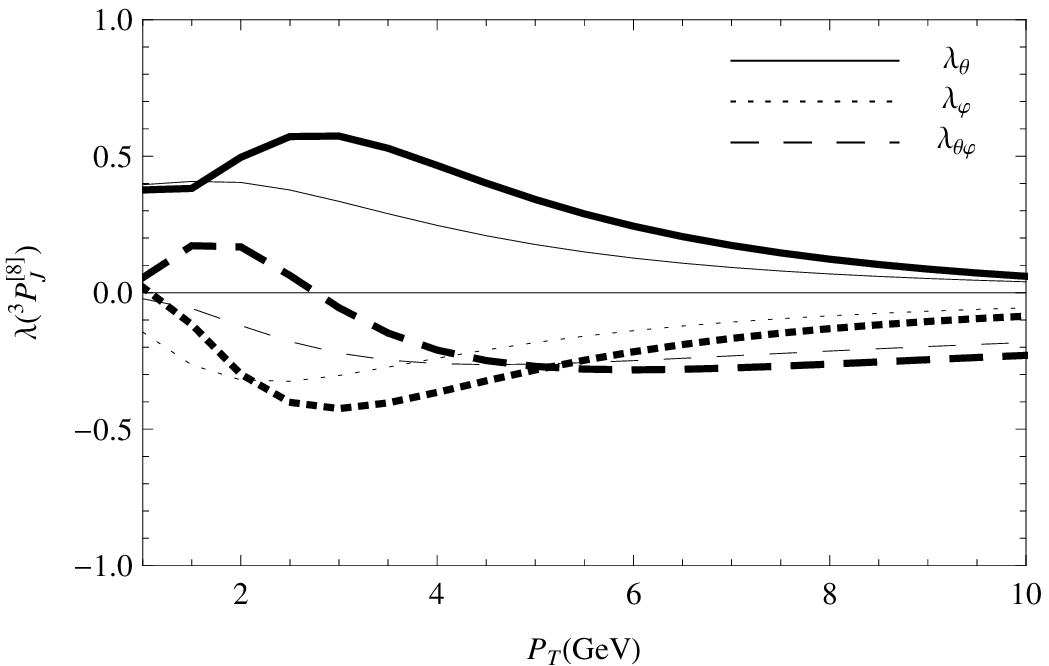}&
\includegraphics[scale=0.80]{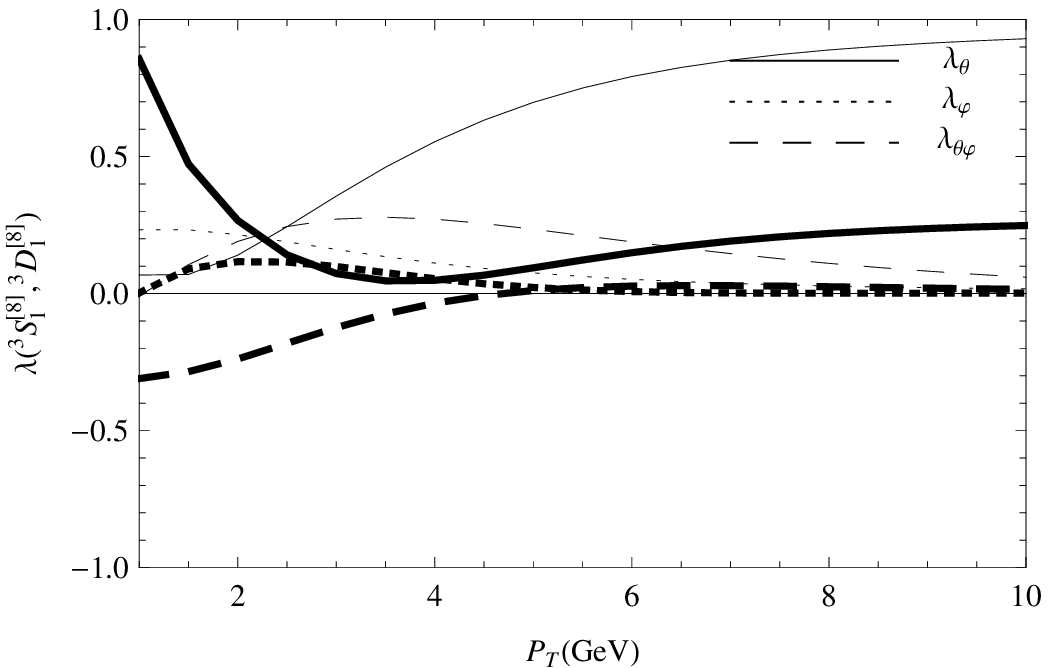}\\
(c)&(d)
\end{tabular}
\caption{$\mathcal{O}(v^2)$-corrected polarization parameters $\lambda_\theta$
(solid lines), $\lambda_\phi$ (dotted lines), and $\lambda_{\theta\phi}$
(dashed lines) in direct $J/\psi$ photoproduction at HERA Run~II
\cite{Aaron:2010gz} through the $c\bar{c}$ Fock states (a) $n={}^3\!S_1^{[1]}$,
(b) ${}^3\!S_1^{[8]}$, (c) ${}^3\!P_J^{[8]}$, and (d)
${}^3\!S_1^{[8]}$--${}^3\!D_1^{[8]}$ mixing as functions of $p_T$.
The true results should lie inside the bands encompassed by the evaluations
with $v^2=0.3$ (thick lines) and the LO results, with $v^2=0$ (thin lines).}
\label{fig:one}
\end{figure}

\begin{figure}
\begin{tabular}{cc}
\includegraphics[scale=0.80]{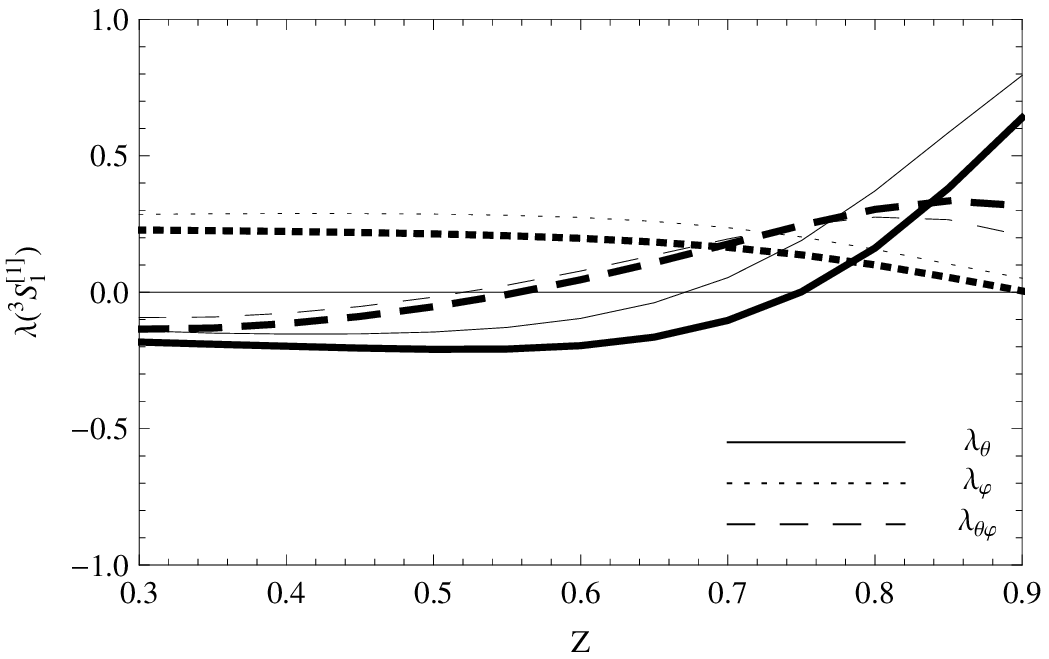}&
\includegraphics[scale=0.80]{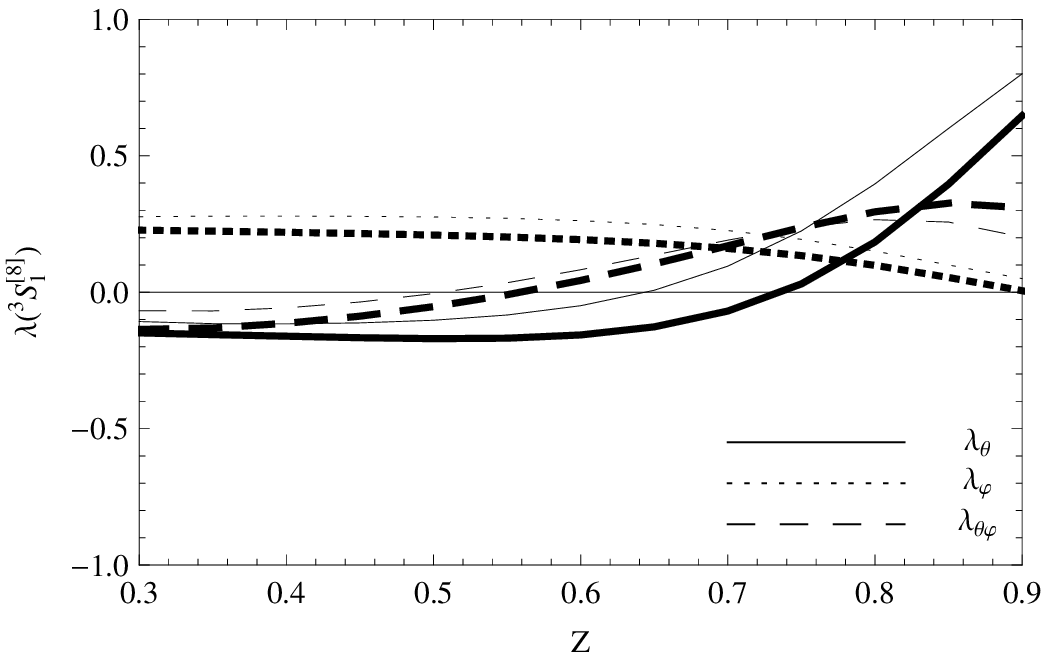}\\
(a)&(b)\\
\includegraphics[scale=0.80]{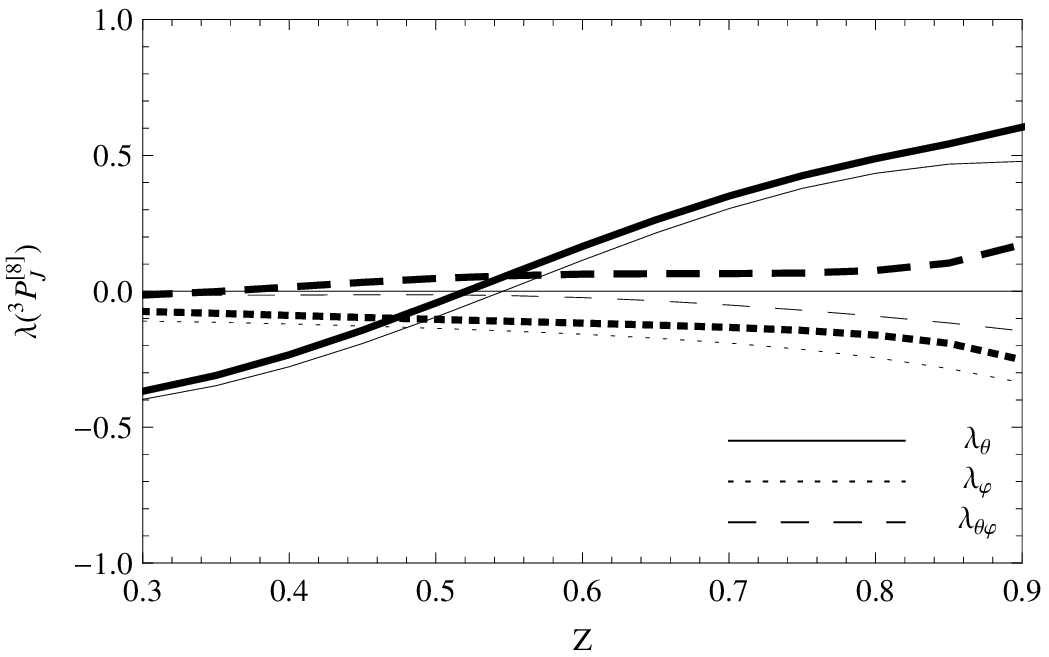}&
\includegraphics[scale=0.80]{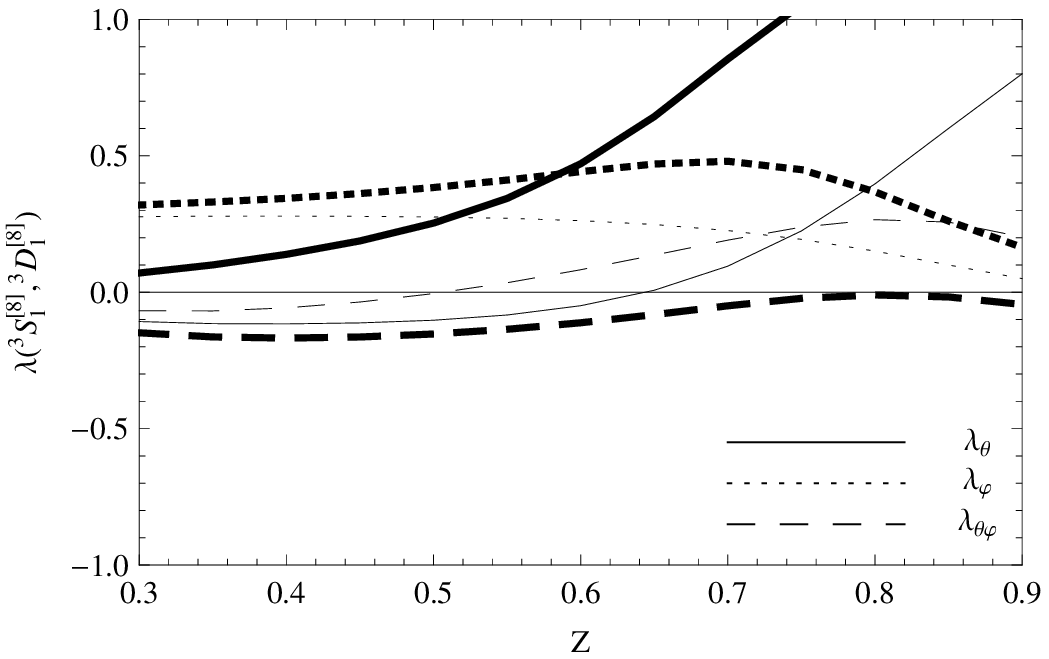}\\
(c)&(d)
\end{tabular}
\caption{Same as in Fig.~\ref{fig:one}, but as functions of $z$.}
\label{fig:two}
\end{figure}

The $J/\psi$ polarization was measured in a number of experiments
\cite{Affolder:2000nn,Abulencia:2007us,Abt:2009nu,Chekanov:2009ad,%
Aaron:2010gz,Adare:2009js,Abelev:2011md,Chatrchyan:2013cla,Aaij:2013nlm}.
As an illustration, we consider here three representative cases, namely
HERA Run~II \cite{Aaron:2010gz}, Tevatron Run~II \cite{Abulencia:2007us}, and
the CMS experimental setup at the LHC \cite{Chatrchyan:2013cla}, limiting
ourselves to the helicity frame.
In HERA Run~II \cite{Aaron:2010gz}, the polarization was measured in prompt
$J/\psi$ photoproduction at center-of-mass (c.m.) energy $\sqrt{S}=319$~GeV
differential in $p_T$ and inelasticity
$z=p_{J/\psi}\cdot p_p/p_\gamma\cdot p_p$, where $p_\gamma$, $p_p$, and
$p_{J/\psi}$ are the four-momenta of the photon, proton, and $J/\psi$ meson,
respectively, imposing in turn the acceptance cuts $0.3<z<0.9$ and
$p_T^2>1~\mathrm{GeV}^2$ always in combination with the acceptance cut
$60~\mathrm{GeV}<W<240~\mathrm{GeV}$ on the $\gamma p$ c.m.\ energy
$W=\sqrt{(p_\gamma+p_p)^2}$ \cite{Aaron:2010gz}.
The $\mathcal{O}(v^2)$ corrections to the polarization parameters
$\lambda_\theta$, $\lambda_\phi$, and $\lambda_{\theta\phi}$ in direct
photoproduction at HERA Run~II \cite{Aaron:2010gz} through the $c\bar{c}$ Fock
states $n={}^3\!S_1^{[1,8]}$, ${}^3\!P_J^{[8]}$, and
${}^3\!S_1^{[8]}$--${}^3\!D_1^{[8]}$ mixing are shown in Figs.~\ref{fig:one}
and \ref{fig:two} as functions of $p_T$ and $z$, respectively.
The true results should lie inside the bands encompassed by the
evaluations with $v^2=0.3$ and the LO results, corresponding to $v^2=0$.
We observe from Fig.~\ref{fig:one} that the $\mathcal{O}(v^2)$ corrections
are generally more significant in the small-$p_T$ range, while they tend to
fade out for asymptotically large values of $p_T$, with the exception of
$\lambda_\phi$ for $n={}^3\!S_1^{[1]}$ and of $\lambda_\theta$ for
${}^3\!S_1^{[8]}$--${}^3\!D_1^{[8]}$ mixing.
Because of the steep falloff of the $p_T$ distributions, the $z$ distributions
in Fig.~\ref{fig:two} receive dominant contributions from the $p_T$ region
close to the cut at 1~GeV, toward the left ends of the panels in
Fig.~\ref{fig:one}.
We observe from Fig.~\ref{fig:two} that the $\mathcal{O}(v^2)$ corrections to
$\lambda_\theta$ and $\lambda_{\theta\phi}$ tend to be more sizable toward
large values of $z$.
The $\mathcal{O}(v^2)$ corrections to $\lambda_\phi$ are relatively modest,
except for the case of ${}^3\!S_1^{[8]}$--${}^3\!D_1^{[8]}$ mixing, where
they are appreciable in the vicinity of $z=0.7$.

\begin{figure}
\begin{tabular}{cc}
\includegraphics[scale=0.80]{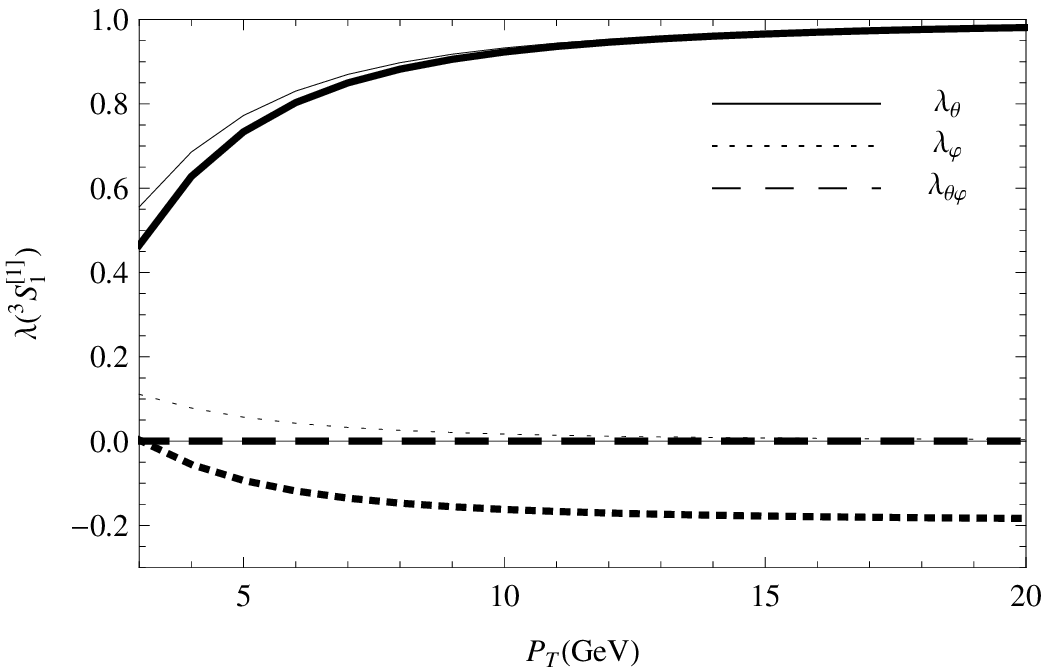}&
\includegraphics[scale=0.80]{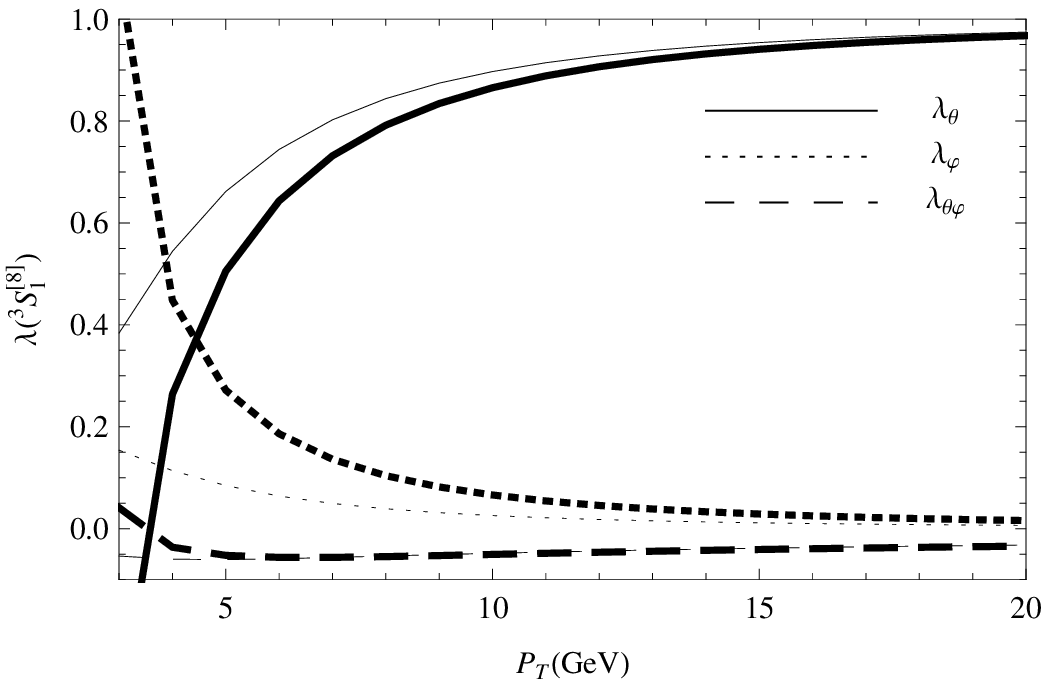}\\
(a)&(b)\\
\includegraphics[scale=0.80]{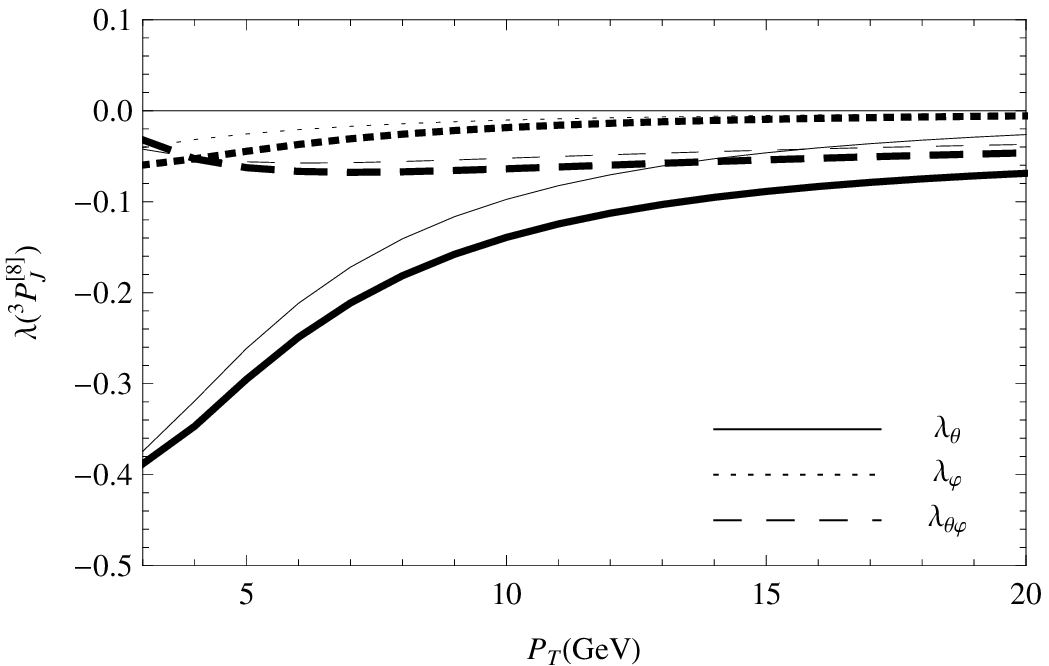}&
\includegraphics[scale=0.80]{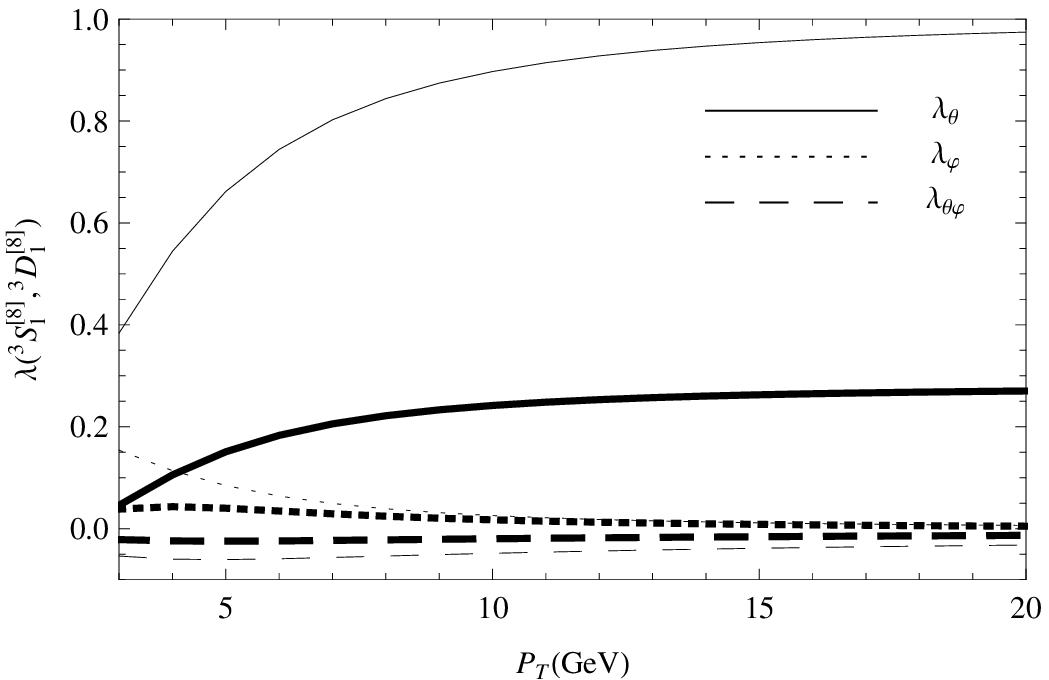}\\
(c)&(d)\\
\includegraphics[scale=0.80]{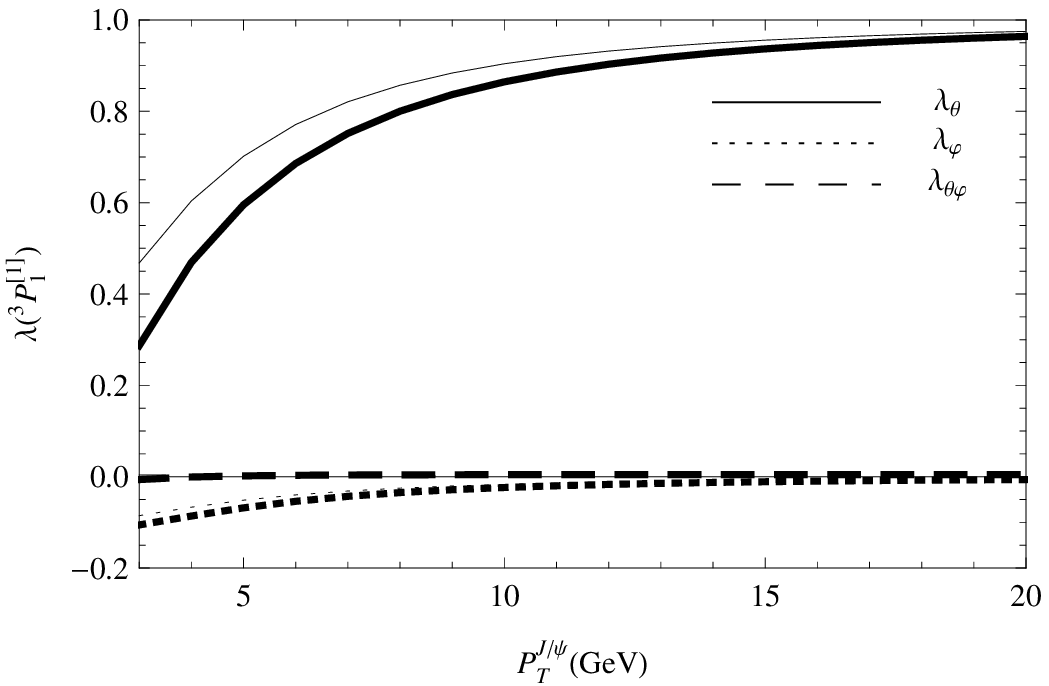}&
\includegraphics[scale=0.80]{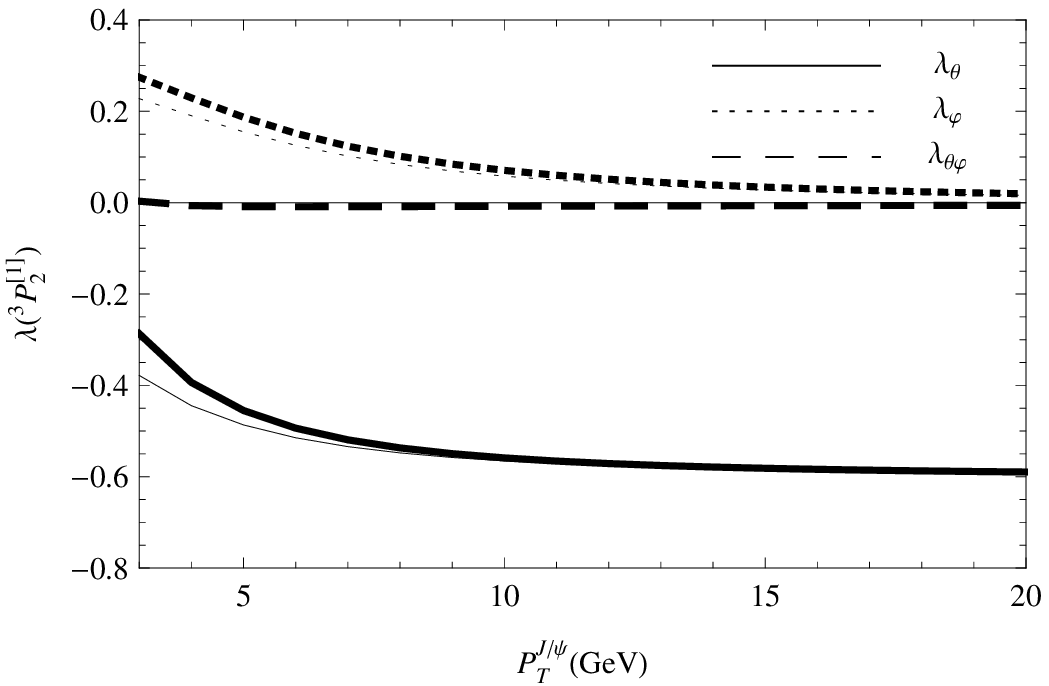}\\
(e)&(f)
\end{tabular}
\caption{$\mathcal{O}(v^2)$-corrected polarization parameters $\lambda_\theta$
(solid lines), $\lambda_\phi$ (dotted lines), and $\lambda_{\theta\phi}$ (dashed
lines) in prompt $J/\psi$ hadroproduction at Tevatron Run~II
\cite{Abulencia:2007us} through the $c\bar{c}$ Fock states (a)
$n={}^3\!S_1^{[1]}$, (b) ${}^3\!S_1^{[8]}$, (c) ${}^3\!P_J^{[8]}$, (d)
${}^3\!S_1^{[8]}$--${}^3\!D_1^{[8]}$ mixing, (e) ${}^3\!P_1^{[1]}$, and (f)
${}^3\!P_2^{[1]}$ as functions of $p_T$.
The true results should lie inside the bands encompassed by the evaluations
with $v^2=0.3$ (thick lines) and the LO results, with $v^2=0$ (thin lines).}
\label{fig:three}
\end{figure}

\begin{figure}
\begin{tabular}{cc}
\includegraphics[scale=0.80]{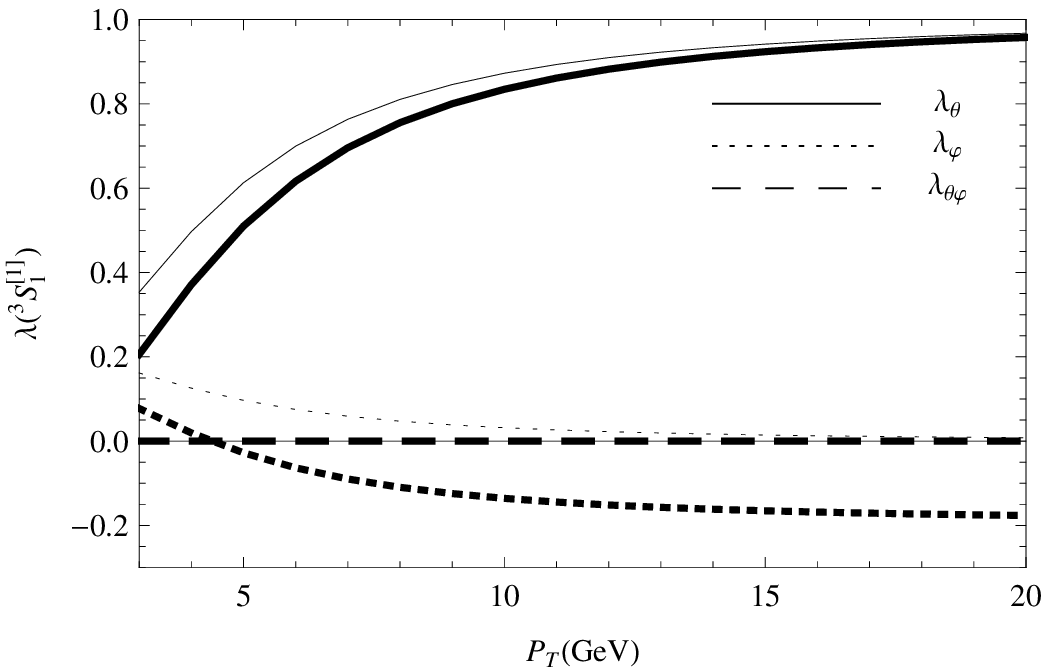}&
\includegraphics[scale=0.80]{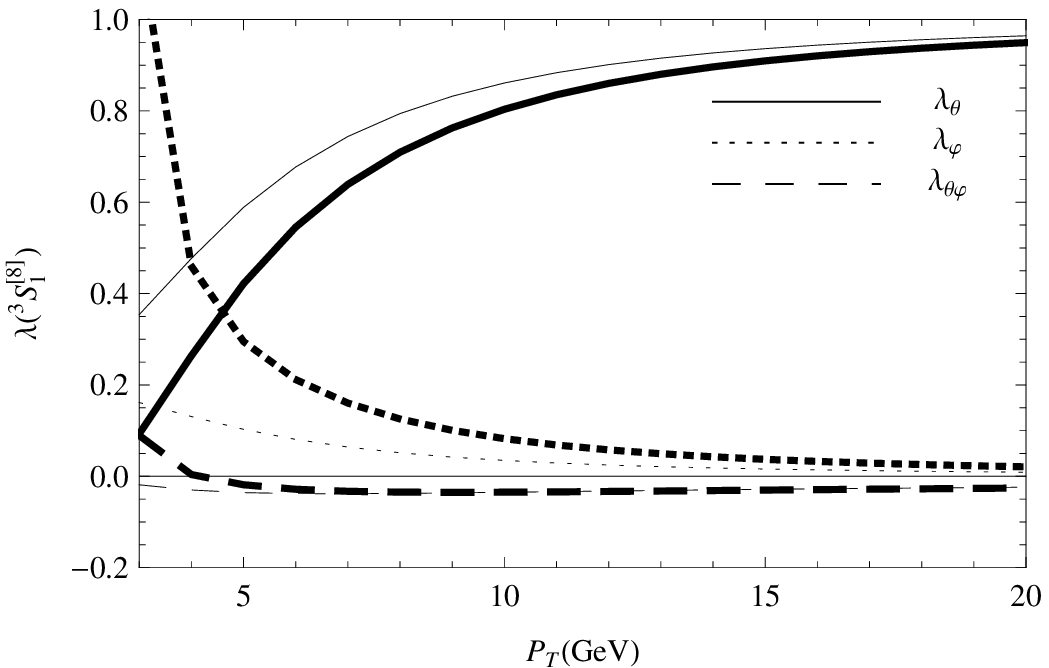}\\
(a)&(b)\\
\includegraphics[scale=0.80]{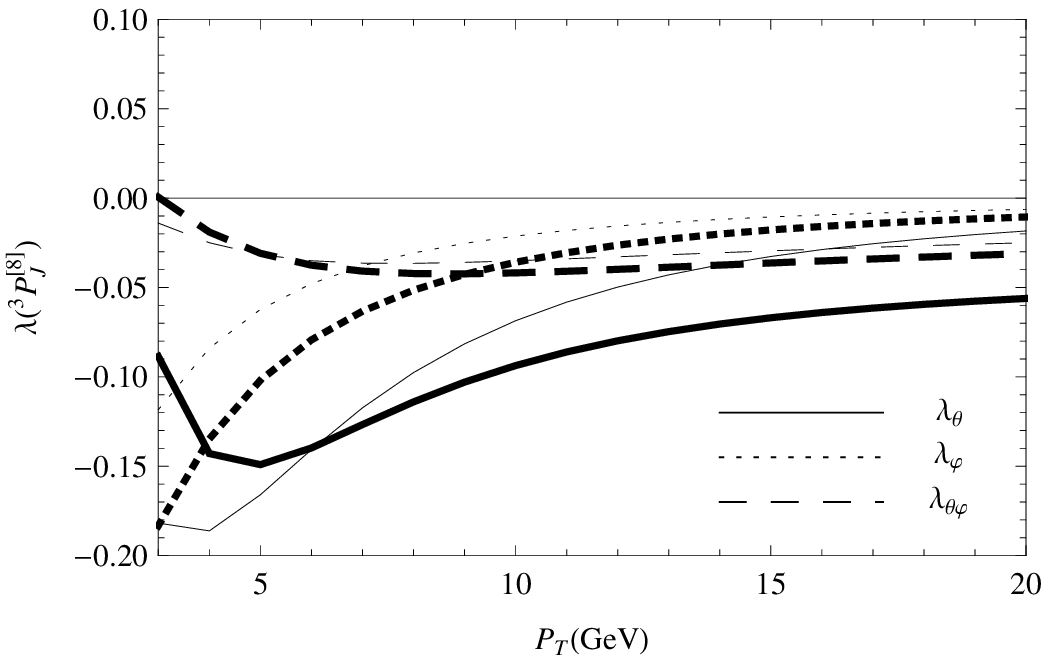}&
\includegraphics[scale=0.80]{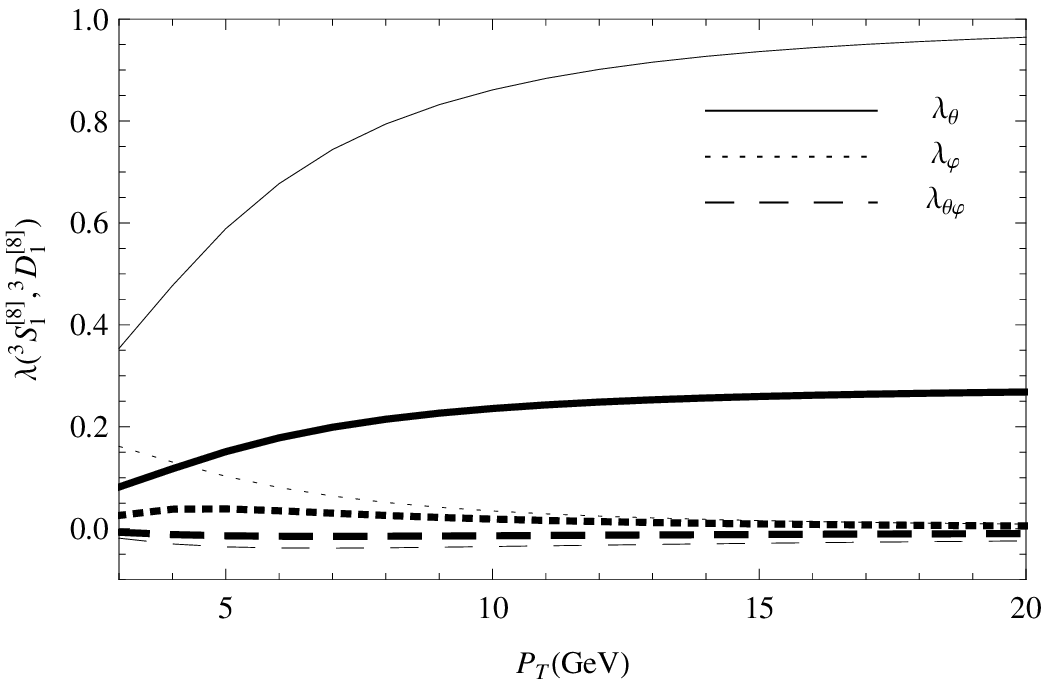}\\
(c)&(d)\\
\includegraphics[scale=0.80]{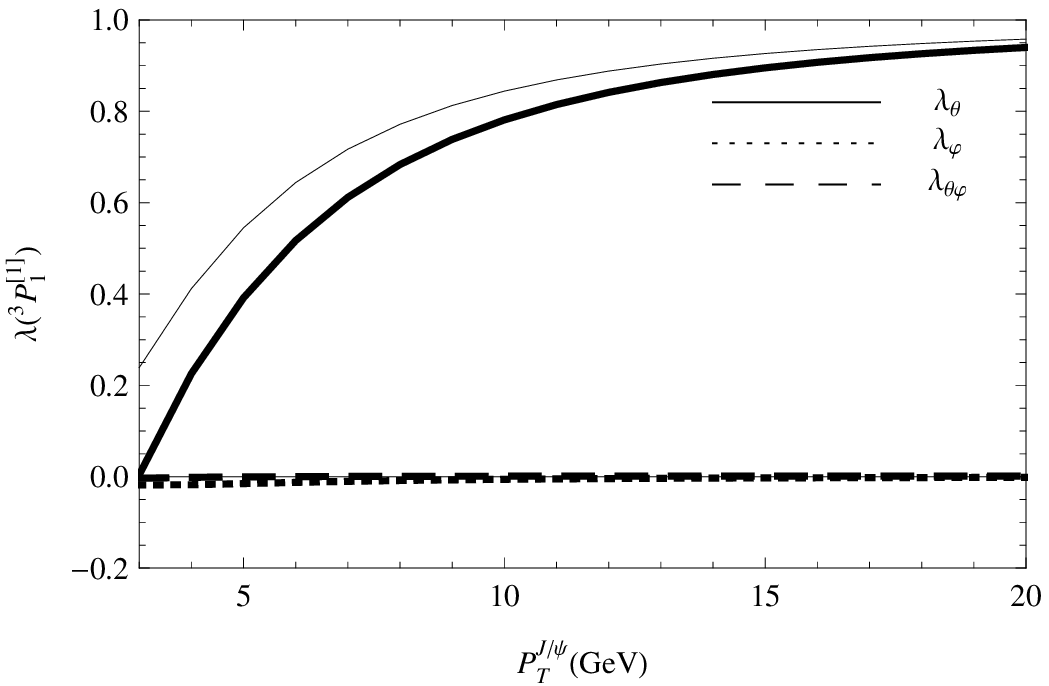}&
\includegraphics[scale=0.80]{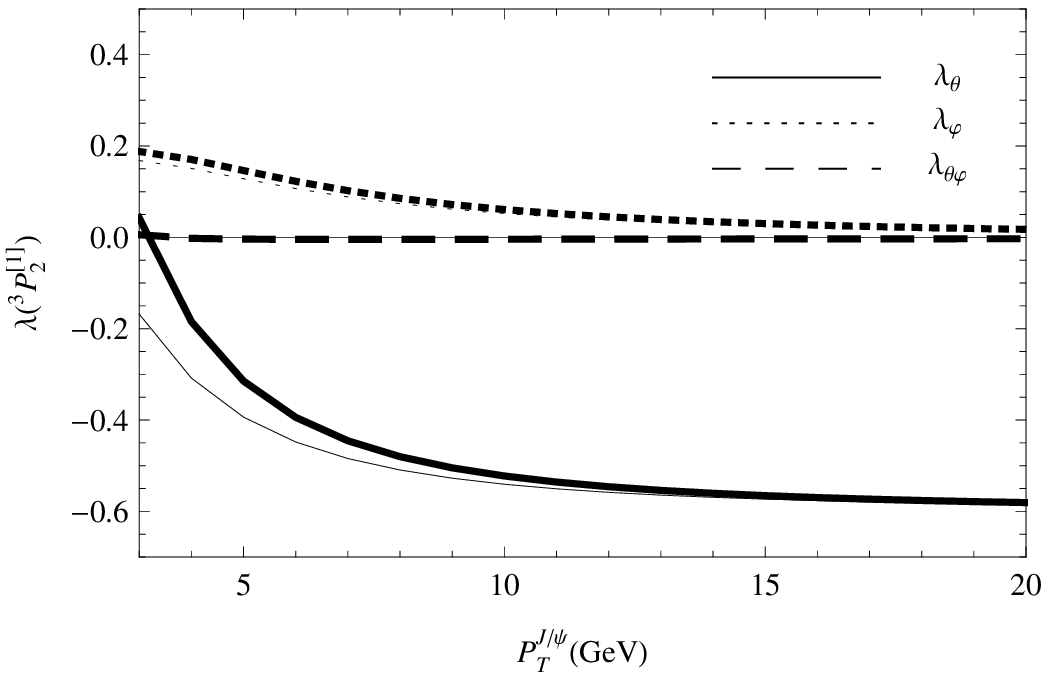}\\
(e)&(f)
\end{tabular}
\caption{Same as in Fig.~\ref{fig:three}, but for the CMS experimental setup at
the LHC \cite{Chatrchyan:2013cla}.}
\label{fig:four}
\end{figure}

The $p_T$ dependence of the polarization in prompt $J/\psi$ hadroproduction was
measured in Tevatron Run~II at $\sqrt{S}=1.96$~TeV for $|y|<0.6$
\cite{Abulencia:2007us} and under CMS experimental conditions at
$\sqrt{S}=7$~TeV for $|y|<2.4$ \cite{Chatrchyan:2013cla}.
In both cases, we exclude from our considerations the small-$p_T$ range,
$p_T<3$~GeV, where the application of fixed-order perturbation theory is
problematic.
The $\mathcal{O}(v^2)$ corrections to the polarization parameters
$\lambda_\theta$, $\lambda_\phi$, and $\lambda_{\theta\phi}$ in prompt
hadroproduction through the $c\bar{c}$ Fock states
$n={}^3\!S_1^{[1,8]}$, ${}^3\!P_J^{[1,8]}$, and
${}^3\!S_1^{[8]}$--${}^3\!D_1^{[8]}$ mixing are shown as functions of $p_T$ for
Tevatron Run~II \cite{Abulencia:2007us} and the CMS setup
\cite{Chatrchyan:2013cla} in Figs.~\ref{fig:three} and \ref{fig:four},
respectively.
We observe from Fig.~\ref{fig:three} that the $\mathcal{O}(v^2)$ corrections to
$\lambda_\theta$ for $n={}^3\!S_1^{[1]}$, ${}^3\!S_1^{[8]}$,
${}^3\!P_{1,2}^{[1]}$ are significant at small values of $p_T$, but quickly
fade out toward large values of $p_T$, while they increase with $p_T$ for
$n={}^3\!P_J^{[8]}$ and ${}^3\!S_1^{[8]}$--${}^3\!D_1^{[8]}$ mixing.
On the other hand, the $\mathcal{O}(v^2)$ corrections to $\lambda_\phi$ are
generally small and fade out as the value of $p_T$ increases, except for the
case of $n={}^3\!S_1^{[1]}$, where the $\mathcal{O}(v^2)$ correction tends
toward a constant negative value for increasing value of $p_T$, while the LO
result tends to zero.
Finally, the $\mathcal{O}(v^2)$ corrections to $\lambda_{\theta\phi}$ are
generally very small for all values of $p_T$.
The situation is very similar for the CMS setup \cite{Chatrchyan:2013cla}, as
may be seen by comparing Figs.~\ref{fig:three} and \ref{fig:four}.

\section{Summary}
\label{sec:four}

In this work, we systematically calculated the relativistic corrections of
relative order $\mathcal{O}(v^2)$ to the polarization observables
$\lambda_\theta$, $\lambda_\phi$, and $\lambda_{\theta\phi}$ in prompt $J/\psi$
photoproduction and hadroproduction, including both the direct mode and the
feed down from $\chi_{cJ}$ and $\psi^\prime$ mesons, and provided in analytic
form the SDCs of all the relevant partonic subprocesses, which are listed in
Eq.~(\ref{eq:sub}).
This study is a natural extension of the one presented in
Ref.~\cite{He:2014sga}, where the $\mathcal{O}(v^2)$ corrections to the
unpolarized $J/\psi$ yields in photoproduction and hadroproduction were
considered.
All these $\mathcal{O}(v^2)$ corrections are to be included along with the
respective $\mathcal{O}(\alpha_s)$ corrections
\cite{Butenschoen:2009zy,Butenschoen:2011ks,Ma:2010vd,Ma:2010yw,%
Butenschoen:2010rq,Butenschoen:2012px,Chao:2012iv,Gong:2012ug,Shao:2014fca,%
Shao:2014yta} to complete the NLO treatments.
We obtained the $\chi_{cJ}$ feed-down contributions in a compact form with the
structure that is familiar from direct production.
We found agreement with the LO results for direct production
\cite{Beneke:1998re} and $\chi_{cJ}$ feed down
\cite{Kniehl:2000nn,Shao:2012fs}.
Our results for the $\mathcal{O}(v^2)$ corrections are new.
Upon summation over the helicity labels, we recover from them our
$\mathcal{O}(v^2)$ results for the unpolarized yield of prompt $J/\psi$
production \cite{He:2014sga}.
We numerically estimated the $\mathcal{O}(v^2)$ corrections to the $J/\psi$
polarization parameters $\lambda_\theta$, $\lambda_\phi$, and
$\lambda_{\theta\phi}$ in direct photoproduction at HERA \cite{Aaron:2010gz}
and prompt hadroproduction at the Tevatron \cite{Abulencia:2007us} and the LHC
\cite{Chatrchyan:2013cla} due to the relevant $c\overline{c}$ Fock states
assuming that their LMDEs obey the velocity scaling rules \cite{Lepage:1992tx}.
We found that the shifts in $\lambda_\theta$, $\lambda_\phi$, and
$\lambda_{\theta\phi}$ due to the $\mathcal{O}(v^2)$ corrections are
significant in most cases, typically reaching $\pm0.2$.
Due to their nontrivial dependencies on $p_T$ and $z$ and the characteristic
differences between photoproduction and hadroproduction, their inclusion in
global data analyses of $J/\psi$ yield and polarization, in addition to the
available $\mathcal{O}(\alpha_s)$ corrections
\cite{Butenschoen:2009zy,Butenschoen:2011ks,Ma:2010vd,Ma:2010yw,%
Butenschoen:2010rq,Butenschoen:2012px,Chao:2012iv,Gong:2012ug,Shao:2014fca,%
Shao:2014yta}, may help to improve the quality of the determinations of the CO
LDMEs and hopefully to remedy the $J/\psi$ polarization crisis.

\section*{Acknowledgments}

This work is supported by the German Federal Ministry for Education and
Research BMBF through Grant No.\ 05H12GUE.

\begin{appendix}
\section{Analytic results}
\label{sec:app}

In this Appendix, we present in compact analytic form the coefficients
$A_k^{ij}(n)$ and $B_k^{ij}(n)$ defined in Eq.~(\ref{rho}) for the partonic
subprocesses in Eq.~(\ref{eq:sub}).
We refrain from presenting $A_k^{ij}(n)$ for
$n={}^3\!S_1^{[1,8]}$, ${}^1\!S_0^{[8]}$, ${}^3\!P_J^{[8]}$ because these
expressions may be found in Appendix~B of Ref.~\cite{Beneke:1998re}.
We list $A^{ij}(n)$ for $n={}^3\!P_{1,2}^{[1]}$ and $B^{ij}(n)$ for
$n={}^3\!S_1^{[1,8]}$, ${}^3\!P_J^{[8]}$, ${}^3\!P_{1,2}^{[1]}$, and
${}^3\!S_1^{[8]}$--${}^3\!D_1^{[8]}$ mixing.
It is convenient to pull out common factors by writing
$A_{k}^{ij}(n)=a^{ij}(n)g_k$ and $B_{k}^{ij}(n)=b^{ij}(n)h_k$.
The Mandelstam variables $s$, $t$, and $u$ appearing below are defined as
$s=(k_1+k_2)^2$, $t=(k_1-P)^2|_{\mathbf{q}=0}$, and
$u=(k_2-P)^2|_{\mathbf{q}=0}$, and satisfy $s+t+u=4m_c^2$.

\subsection{Photoproduction}

$g+\gamma\to c\bar{c}({}^3\!S_1^{[1]})+g$:

\begin{subequations}
\begin{eqnarray}
b^{g\gamma}({}^3\!S_1^{[1]})
=\frac{4096 \pi ^3 \alpha  \alpha_s^2 }{729 m_c \left(4 m_c^2-s\right)^3
\left(4 m_c^2-t\right)^3 (s+t)^3}
\end{eqnarray}
\begin{eqnarray}
h_1&=&2048 m_c^{10} \left(3 s^2+2 s t+3 t^2\right)-256 m_c^8
\left(5 s^3-12 s^2 t+5 t^3\right)-64 m_c^6 (15 s^4+80 s^3 t\nonumber\\
&+&90 s^2 t^2+44 s t^3+15 t^4 )+16 m_c^4 \left(21 s^5+103 s^4 t
+182 s^3 t^2+158 s^2 t^3+67 s t^4+21 t^5\right)\nonumber\\
&-& 4 m_c^2 \left(7 s^6+42 s^5 t+99 s^4 t^2+124 s^3 t^3+87 s^2 t^4
+30 s t^5+7 t^6\right)\nonumber\\
&+&3 s t (s+t) \left(s^2+s t+t^2\right)^2
\end{eqnarray}
\begin{eqnarray}
h_2&=&8 \Big{[}4096 m_c^{12}-768 m_c^{10} (s-3 t)+64 m_c^8
\left(5 s^2-50 s t-43 t^2\right)+16 m_c^6 ( 8 s^3+67 s^2 t \nonumber\\
&+&98  s t^2+47 t^3 )+4 m_c^4 \left(2 s^4-20 s^3 t-59 s^2 t^2-58 s t^3
-25 t^4\right)+m_c^2 t (-6 s^4-4 s^3 t\nonumber\\
&+&15 s^2 t^2+21 s t^3+12 t^4)-t^2 \left(s^2+s t+t^2\right)^2\Big{]}
\end{eqnarray}
\begin{eqnarray}
h_3&=&8 \Big{[}512 m_c^8 \left(s^2+t^2\right)+16 m_c^6 \left(13 s^3
+33 s^2 t+9 s t^2+13 t^3\right)-4 m_c^4 (13 s^4+58 s^3 t\nonumber\\
&+&74 s^2 t^2+46 s t^3+25 t^4)+m_c^2 (s+t) \left(12 s^4+29 s^3 t
+48 s^2 t^2+29 s t^3+12 t^4\right)\nonumber\\
&-&(s+t)^2 \left(s^2+s t+t^2\right)^2\Big{]}
\end{eqnarray}
\begin{eqnarray}
h_4&=&8 \Big{[}128 m_c^8 \left(5 s^2-t^2\right)+16 m_c^6 \left(5 s^3
+19 s^2 t+14 s t^2+8 t^3\right)+4 m_c^4 (s^4-22 s^3 t\nonumber\\
&-&49 s^2 t^2-36 s t^3-14 t^4)+m_c^2 t \left(7 s^4+29 s^3 t+48 s^2 t^2
+34 s t^3+12 t^4\right)\nonumber\\
&-&t (s+t) \left(s^2+s t+t^2\right)^2\Big{]}
\end{eqnarray}
\label{eq:ten}
\end{subequations}

$g+\gamma\to c\bar{c}({}^3S_1^{[8]})+g$:
\begin{eqnarray}
b^{g\gamma}({}^3\!S_1^{[8]})=\frac{15}{8}b^{g\gamma}({}^3\!S_1^{[1]}).
\end{eqnarray}
The coefficients $h_k$ are the same as those in Eq.~(\ref{eq:ten}).

$g+\gamma\to c\bar{c}({}^3\!P_J^{[8]})+g$:
\begin{subequations}
\begin{eqnarray}
b^{g\gamma}({}^3\!P_J^{[8]})=\frac{128 \pi ^3 \alpha  \alpha_s^2}{15 m_c^5 \left(s-4 m_c^2\right)^4
\left(t-4 m_c^2\right)^4 (s+t)^4 \left(-4 m_c^2+s+t\right)^2}
\end{eqnarray}
\begin{eqnarray}
h_1&=&\left(-4 m_c^2+s+t\right) \Big{[}-524288 m_c^{18} (s+t) \left(3 s^2
+14 s t+3 t^2\right)+65536 m_c^{16} (3 s-t) (3 s^3\nonumber\\
&-&3 s^2 t+25 s t^2-9 t^3)+16384 m_c^{14} \left(39 s^5+421 s^4 t
+568 s^3 t^2+908 s^2 t^3+521 s t^4+39 t^5\right)\nonumber\\
&-&4096 m_c^{12} \left(126 s^6+1064 s^5 t+2089 s^4 t^2+3138 s^3 t^3
+2909 s^2 t^4+1064 s t^5+126 t^6\right)\nonumber\\
&+&1024 m_c^{10} (144 s^7+1050 s^6 t+2473 s^5 t^2+4177 s^4 t^3
+4857 s^3 t^4+3273 s^2 t^5+930 s t^6\nonumber\\
&+&144 t^7)-256 m_c^8 (75 s^8+349 s^7 t+816 s^6 t^2
+1732 s^5 t^3+2564 s^4 t^4+2432 s^3 t^5\nonumber\\
&+&1376 s^2 t^6+249 s t^7+75 t^8)+64 m_c^6 (s+t)
(15 s^8-66 s^7 t-314 s^6 t^2-374 s^5 t^3\nonumber\\
&-&230 s^4 t^4-254 s^3 t^5+26 s^2 t^6-106 s t^7+15 t^8)
+16 m_c^4 s t (s+t) (39 s^7+148 s^6 t\nonumber\\
&+&219 s^5 t^2+193 s^4 t^3+213 s^3 t^4+199 s^2 t^5+88 s t^6+39 t^7)
+4 m_c^2 s^2 t^2 (s+t) (13 s^6 +90 s^5 t\nonumber\\
&+&232 s^4 t^2+286 s^3 t^3+212 s^2 t^4+70 s t^5+13 t^6)
-11 s^3 t^3 (s+t)^2 \left(s^2+s t+t^2\right)^2\Big{]}
\end{eqnarray}
\begin{eqnarray}
h_2&=&-16 m_c^2 \Big{[}-2097152 m_c^{20} (s+t)+131072 m_c^{18}
 \left(31 s^2-70 s t-5 t^2\right)\nonumber\\
&-&65536 m_c^{16} \left(18 s^3-191 s^2 t-180 s t^2-61 t^3\right)
-8192 m_c^{14} (26 s^4+1098 s^3 t+1287 s^2 t^2\nonumber\\
&+&776 s t^3+349 t^4)+4096 m_c^{12} \left(42 s^5+775 s^4 t
+1084 s^3 t^2+556 s^2 t^3+238 s t^4+161 t^5\right)\nonumber\\
&-&512 m_c^{10} \left(79 s^6+972 s^5 t+1074 s^4 t^2
-584 s^3 t^3-1622 s^2 t^4-1350 s t^5-209 t^6\right)\nonumber\\
&+&256 m_c^8 \left(16 s^7+19 s^6 t-602 s^5 t^2-1972 s^4 t^3
-2788 s^3 t^4-2582 s^2 t^5-1612 s t^6-365 t^7\right)\nonumber\\
&-&32 m_c^6 (2 s^8-246 s^7 t-1554 s^6 t^2-4296 s^5 t^3
-6520 s^4 t^4-6780 s^3 t^5-5509 s^2 t^6\nonumber\\
&-&3008 s t^7-673 t^8)-16 m_c^4 t (36 s^8+205 s^7 t
+664 s^6 t^2+1260 s^5 t^3+1731 s^4 t^4+1755 s^3 t^5\nonumber\\
&+&1412 s^2 t^6+702 s t^7+143 t^8)+2 m_c^2 t^2 (-26 s^8
-54 s^7 t+5 s^6 t^2+228 s^5 t^3+648 s^4 t^4\nonumber\\
&+&790 s^3 t^5+669 s^2 t^6+296 s t^7+48 t^8)-s t^3 (s+t)
(6 s^6+21 s^5 t+34 s^4 t^2+42 s^3 t^3\nonumber\\
&+&32 s^2 t^4+19 s t^5+6 t^6)\Big{]}
\end{eqnarray}
\begin{eqnarray}
h_3&=&-16 m_c^2 \Big{[}262144 m_c^{16} (s+t) \left(s^2+t^2\right)
-8192 m_c^{14} (21 s^4+218 s^3 t+114 s^2 t^2+258 s t^3\nonumber\\
&+&21 t^4)+4096 m_c^{12} \left(9 s^5+210 s^4 t+328 s^3 t^2
+648 s^2 t^3+295 s t^4+14 t^5\right)\nonumber\\
&+&2048 m_c^{10} \left(34 s^6+72 s^5 t+50 s^4 t^2-359 s^3 t^3
-490 s^2 t^4-33 s t^5+14 t^6\right)\nonumber\\
&-&256 m_c^8 (s+t) \left(218 s^6+821 s^5 t+1234 s^4 t^2+605 s^3 t^3
-31 s^2 t^4+486 s t^5+143 t^6\right)\nonumber\\
&+&32 m_c^6 (s+t) (497 s^7+2427 s^6 t+4644 s^5 t^2
+5088 s^4 t^3+3648 s^3 t^4+2494 s^2 t^5\nonumber\\
&+&1697 s t^6+397 t^7)-16 m_c^4 (s+t) (125 s^8
+745 s^7 t+1732 s^6 t^2+2347 s^5 t^3+2260 s^4 t^4\nonumber\\
&+&1767 s^3 t^5+1162 s^2 t^6+575 s t^7+115 t^8)+2 m_c^2
(s+t)^2 (48 s^8+312 s^7 t+725 s^6 t^2\nonumber\\
&+&970 s^5 t^3+938 s^4 t^4+770 s^3 t^5+505 s^2 t^6+252 s t^7+48 t^8)
-s t (s+t)^3 (6 s^6+21 s^5 t\nonumber\\
&+&38 s^4 t^2+48 s^3 t^3+38 s^2 t^4+21 s t^5+6 t^6)\Big{]}
\end{eqnarray}
\begin{eqnarray}
h_4&=&-16 m_c^2 \Big{[}1048576 m_c^{18} (s-t) (s+t)
-98304 m_c^{16} \left(5 s^3-11 s t^2-6 t^3\right)\nonumber\\
&+&8192 m_c^{14} \left(s^4-255 s^3 t-197 s^2 t^2-93 s t^3
+8 t^4\right)+2048 m_c^{12} (90 s^5+897 s^4 t\nonumber\\
&+&1281 s^3 t^2+645 s^2 t^3-87 s t^4-122 t^5)-512 m_c^{10}
(226 s^6+1467 s^5 t+2568 s^4 t^2\nonumber\\
&+&1908 s^3 t^3+262 s^2 t^4-895 s t^5-372 t^6)+128 m_c^8
(223 s^7+1197 s^6 t+2055 s^5 t^2\nonumber\\
&+&1372 s^4 t^3-390 s^3 t^4-1905 s^2 t^5-2010 s t^6-614 t^7)
-32 m_c^6 (99 s^8+404 s^7 t+282 s^6 t^2\nonumber\\
&-&989 s^5 t^3-2828 s^4 t^4-3803 s^3 t^5-3693 s^2 t^6-2248 s t^7
-556 t^8)+8 m_c^4 (18 s^9+16 s^8 t\nonumber\\
&-&385 s^7 t^2-1503 s^6 t^3-2951 s^5 t^4-3933 s^4 t^5-3790 s^3 t^6
-2794 s^2 t^7-1296 s t^8-258 t^9)\nonumber\\
&+&2 m_c^2 t (s+t) (16 s^8+145 s^7 t+409 s^6 t^2+681 s^5 t^3
+877 s^4 t^4+836 s^3 t^5+602 s^2 t^6\nonumber\\
&+&274 s t^7+48 t^8)-s t^2 (s+t)^2 \left(7 s^6+23 s^5 t+39 s^4 t^2
+47 s^3 t^3+36 s^2 t^4+20 s t^5+6 t^6\right)\Big{]}\nonumber\\
\end{eqnarray}
\end{subequations}

$g+\gamma\to c\bar{c}({}^3\!S_1^{[8]},{}^3\!D_1^{[8]})+g$:
\begin{subequations}
\begin{eqnarray}
b^{g\gamma}({}^3\!S_1^{[8]},{}^3\!D_1^{[8]})
=\frac{1024 \sqrt{\frac{5}{3}} \pi ^3 \alpha  \alpha_s^2 }
{81 m_c \left(s-4 m_c^2\right)^4 \left(t-4 m_c^2\right)^4 (s+t)^4}
\end{eqnarray}
\begin{eqnarray}
h_1&=&-\left(4 m_c^2-s\right) \left(4 m_c^2-t\right) (s+t)\Big{[}
2048 m_c^{10} \left(3 s^2+5 s t+3 t^2\right)-256 m_c^8 (5 s^3+24 s^2 t\nonumber\\
&+&24 s t^2+5 t^3)-64 m_c^6 \left(15 s^4+2 s^3 t-18 s^2 t^2+2 s t^3+
15 t^4\right)+16 m_c^4 (21 s^5+31 s^4 t\nonumber\\
&+&20 s^3 t^2+20 s^2 t^3+31 s t^4+21 t^5)-4 m_c^2 (7 s^6+12 s^5 t+9 s^4 t^2
-2 s^3 t^3+9 s^2 t^4+12 s t^5\nonumber\\
&+&7 t^6)-3 s t (s+t) \left(s^2+s t+t^2\right)^2\Big{]}
\end{eqnarray}
\begin{eqnarray}
h_2&=&-4 \left(4 m_c^2-t\right) \Big{[}81920 m_c^{14} (s+t)-2048
m_c^{12} \left(25 s^2+52 s t+15 t^2\right)\nonumber\\
&+&512 m_c^{10} \left(26 s^3+75 s^2 t+36 s t^2-13 t^3\right)
-128 m_c^8 (3 s^4+18 s^3 t-24 s^2 t^2-92 s t^3\nonumber\\
&-&41 t^4)-32 m_c^6 \left(12 s^5+39 s^4 t+100 s^3 t^2+184 s^2 t^3
+166 s t^4+43 t^5\right)+8 m_c^4 (4 s^6\nonumber\\
&+&30 s^5 t+81 s^4 t^2+140 s^3 t^3+161 s^2 t^4+106 s t^5+30 t^6)
-2 m_c^2 t (6 s^6+12 s^5 t+29 s^4 t^2\nonumber\\
&+&38 s^3 t^3+41 s^2 t^4+24 s t^5+10 t^6)-s t^2 (s+t)
\left(s^2+s t+t^2\right)^2\Big{]}
\end{eqnarray}
\begin{eqnarray}
h_3&=&-4 (s+t) \Big{[}4096 m_c^{12} \left(s^2+t^2\right)+512
m_c^{10} \left(29 s^3+61 s^2 t+61 s t^2+29 t^3\right)\nonumber\\
&-&256 m_c^8 (s+t)^2 \left(37 s^2+43 s t+37 t^2\right)+32 m_c^6
(73 s^5+299 s^4 t+536 s^3 t^2+536 s^2 t^3\nonumber\\
&+&299 s t^4+73 t^5)-16 m_c^4 \left(20 s^6+80 s^5 t+167 s^4 t^2
+212 s^3 t^3+167 s^2 t^4+80 s t^5+20 t^6\right)\nonumber\\
&+&2 m_c^2 \left(10 s^7+32 s^6 t+59 s^5 t^2+77 s^4 t^3+77 s^3 t^4
+59 s^2 t^5+32 s t^6+10 t^7\right)+s t (s^3\nonumber\\
&+&2 s^2 t+2 s t^2+t^3)^2\Big{]}
\end{eqnarray}
\begin{eqnarray}
h_4&=&-4 \Big{[}2048 m_c^{12} \left(13 s^3+7 s^2 t-5 s t^2-11 t^3\right)
-512 m_c^{10} (14 s^4+5 s^3 t-61 s^2 t^2-95 s t^3\nonumber\\
&-&43 t^4)-128 m_c^8 \left(s^5+43 s^4 t+214 s^3 t^2+340 s^2 t^3+265 s t^4
+73 t^5\right)+32 m_c^6 (2 s^6\nonumber\\
&+&71 s^5 t+305 s^4 t^2+536 s^3 t^3+530 s^2 t^4+301 s t^5+71 t^6)
-8 m_c^4 t (40 s^6+167 s^5 t\nonumber\\
&+&345 s^4 t^2+413 s^3 t^3+327 s^2 t^4+160 s t^5+40 t^6)+2 m_c^2 t
(10 s^7+30 s^6 t+57 s^5 t^2\nonumber\\
&+&77 s^4 t^3+79 s^3 t^4+61 s^2 t^5+32 s t^6+10 t^7)+s t^2 \left(s^3+2 s^2
t+2 s t^2+t^3\right)^2\Big{]}
\end{eqnarray}
\end{subequations}

$q(\bar{q})+\gamma\to c\bar{c}({}^3\!S_1^{[8]})+q(\bar{q})$:
\begin{subequations}
\begin{eqnarray}
b^{q(\bar{q})\gamma}({}^3\!S_1^{[8]})
=\frac{4 \pi ^3 \alpha  \alpha_s^2 e_q^2}{27 m_c^5 s t \left(4 m_c^2-s\right)}
\end{eqnarray}
\begin{eqnarray}
h_1=-640 m_c^6+160 m_c^4 (2 s+t)-4 m_c^2 \left(27 s^2+10 s t+5 t^2\right)+11 s \left(s^2+t^2\right)
\end{eqnarray}
\begin{eqnarray}
h_2=-16 m_c^2 \left(44 m_c^2-5 s\right)
\end{eqnarray}
\begin{eqnarray}
h_3=-32 m_c^2 \left(44 m_c^2-5 s\right)
\end{eqnarray}
\begin{eqnarray}
h_4=-16 m_c^2 \left(44 m_c^2-5 s\right)
\end{eqnarray}
\end{subequations}

$q(\bar{q})+\gamma\to c\bar{c}({}^3\!P_J^{[8]})+q(\bar{q})$:
\begin{subequations}
\begin{eqnarray}
b^{q(\bar{q})\gamma}({}^3\!P_J^{[8]})=\frac{256 \pi ^3 \alpha  
\alpha_s^2}{135 m_c^5 \left(4 m_c^2-s\right)
(s+t)^4 \left(-4 m_c^2+s+t\right)^2}
\end{eqnarray}
\begin{eqnarray}
h_1&=&(s+t) \left(s+t-4 m_c^2\right) \Big{[}1280 m_c^6 (s+5 t)
-128 m_c^4 t (6 s+5 t)+4 m_c^2 (21 s^3\nonumber\\
&+&49 s^2 t+s t^2+21 t^3)-11 s \left(s^3+s^2 t+s t^2+t^3\right)\Big{]}
\end{eqnarray}
\begin{eqnarray}
h_2&=&-32 m_c^2 \Big{[}1024 m_c^8-128 m_c^6 (s-14 t)+16 m_c^4
\left(21 s^2+14 s t-29 t^2\right)-4 m_c^2 (4 s^3\nonumber\\
&+&13 s^2 t+54 s t^2+3 t^3)+s \left(s^3+s^2 t+3 s t^2+3 t^3\right)\Big{]}
\end{eqnarray}
\begin{eqnarray}
h_3=-64 m_c^2 (s+t) \Big{[}80 m_c^4 (3 s+7 t)-8 m_c^2
\left(3 s^2+11 s t+8 t^2\right)+s (s+t)^2\Big{]}
\end{eqnarray}
\begin{eqnarray}
h_4&=&-64 m_c^2 \Big{[}64 m_c^6 (s-t)+8 m_c^4 \left(19 s^2
+72 s t+19 t^2\right)-2 m_c^2 (s^3+27 s^2 t+43 s t^2\nonumber\\
&+&17 t^3)+s t (s+t)^2\Big{]}
\end{eqnarray}
\label{eq:fifteen}
\end{subequations}

$q(\bar{q})+\gamma\to c\bar{c}({}^3\!S_1^{[8]},{}^3\!D_1^{[8]})+q(\bar{q})$:
\begin{subequations}
\begin{eqnarray}
b^{q(\bar{q})\gamma}({}^3\!S_1^{[8]},^3D_1^{[8]})=\frac{8 \sqrt{\frac{5}{3}} 
\pi ^3 \alpha  \alpha_s^2 e_q^2}{9 m_c^5 s t}
\end{eqnarray}
\begin{eqnarray}
h_1=32 m_c^4-8 m_c^2 (s+t)+s^2+t^2
\end{eqnarray}
\begin{eqnarray}
h_2=16 m_c^2
\end{eqnarray}
\begin{eqnarray}
h_3=32 m_c^2
\end{eqnarray}
\begin{eqnarray}
h_4=16 m_c^2
\end{eqnarray}
\end{subequations}

\subsection{Hadroproduction}

$g+g\to c\bar{c}({}^3\!S_1^{[1]})+g$:
\begin{eqnarray}
b^{gg}({}^3\!S_1^{[1]})
=\frac{15\alpha_s}{128\alpha}b^{g\gamma}({}^3\!S_1^{[1]}).
\end{eqnarray}
The coefficients $h_k$ are the same as those in Eq.~(\ref{eq:ten}).

$g+g\to c\bar{c}({}^3\!P_1^{[1]})+g$:
\begin{subequations}
\begin{eqnarray}
a^{gg}({}^3\!P_1^{[1]})
=\frac{16 \pi ^3 \alpha_s^3}{9 m_c^3 \left(s-4 m_c^2\right)^4
\left(t-4 m_c^2\right)^4 (s+t)^4}
\end{eqnarray}
\begin{eqnarray}
g_1&=&\left(s-4 m_c^2\right) \left(4 m_c^2-t\right) (s+t)
\big{[}16 m_c^4 \left(s^2+s t+t^2\right)-4 m_c^2
(2 s^3+3 s^2 t+3 s t^2\nonumber\\
&+&2 t^3)+\left(s^2+s t+t^2\right)^2\big{]} \Big{[}16 m_c^6 (s+t)
-4 m_c^4 \left(5 s^2+6 s t+5 t^2\right)+m_c^2
(8 s^3+13 s^2 t\nonumber\\
&+&13 s t^2+8 t^3)-\left(s^2+s t+t^2\right)^2\Big{]}
\end{eqnarray}
\begin{eqnarray}
g_2&=&-4 m_c^2\left( t-4 m_c^2\right)^2 \Big{[}512 m_c^{10}
(s+t)^2-768 m_c^8 (s+t)^3+32 m_c^6 (13 s^4+56 s^3 t\nonumber\\
&+&86 s^2 t^2+58 s t^3+15 t^4)-16 m_c^4 \left(6 s^5+40 s^4 t
+82 s^3 t^2+85 s^2 t^3+45 s t^4+10 t^5\right)\nonumber\\
&+&2 m_c^2 \left(4 s^6+54 s^5 t+137 s^4 t^2+192 s^3 t^3+153 s^2 t^4
+70 s t^5+14 t^6\right)\nonumber\\
&-&t \left(s^2+s t+t^2\right)^2 \left(7 s^2+7 s t+2 t^2\right)\Big{]}
\end{eqnarray}
\begin{eqnarray}
g_3&=&4 m_c^2 (s+t)^2 \Big{[}512 m_c^{10} \left(s^2+t^2\right)
-256 m_c^8 \left(3 s^3+2 s^2 t+2 s t^2+3 t^3\right)\nonumber\\
&+&32 m_c^6 \left(15 s^4+18 s^3 t+10 s^2 t^2+18 s t^3+15 t^4\right)
-16 m_c^4 (10 s^5+15 s^4 t+9 s^3 t^2+9 s^2 t^3\nonumber\\
&+&15 s t^4+10 t^5)+2 m_c^2 \left(14 s^6+22 s^5 t+17 s^4 t^2+8 s^3 t^3
+17 s^2 t^4+22 s t^5+14 t^6\right)\nonumber\\
&-&\left(s^2+s t+t^2\right)^2 \left(2 s^3-s^2 t-s t^2+2 t^3\right)\Big{]}
\end{eqnarray}
\begin{eqnarray}
g_4&=&4 m_c^2 \left(s-4 m_c^2\right) \Big{[}512 m_c^{10}
(s-t) (s+t)^2-128 m_c^8 (7 s^4+14 s^3 t+2 s^2 t^2-12 s t^3\nonumber\\
&-&7 t^4)+32 m_c^6 \left(18 s^5+53 s^4 t+46 s^3 t^2-18 s^2 t^3-44 s t^4
-19 t^5\right)-8 m_c^4 (22 s^6+86 s^5 t\nonumber\\
&+&119 s^4 t^2+46 s^3 t^3-60 s^2 t^4-72 s t^5-25 t^6)+2 m_c^2 (14 s^7
+64 s^6 t+114 s^5 t^2+95 s^4 t^3\nonumber\\
&+&3 s^3 t^4-63 s^2 t^5-55 s t^6-16 t^7)-\left(s^2+s t+t^2\right)^2 \left(2 s^4
+5 s^3 t+3 s^2 t^2-4 s t^3-2 t^4\right)\Big{]}\nonumber\\
\end{eqnarray}
\begin{eqnarray}
b^{gg}({}^3\!P_1^{[1]})
=\frac{8 \pi ^3 \alpha_s^3}{45 m_c^5 \left(4 m_c^2-s\right)^5
\left(4 m_c^2-t\right)^5 (s+t)^5}
\end{eqnarray}
\begin{eqnarray}
h_1&=&\left(4 m_c^2-s\right) \left(4 m_c^2-t\right) (s+t) \Big{[}65536
m_c^{16} \left(s^3+s^2 t+s t^2+t^3\right)-4096 m_c^{14} (7 s^4 \nonumber\\
&-&73 s^3 t-60 s^2 t^2-53 s t^3+7 t^4)-1024 m_c^{12} (132 s^5
+853 s^4 t+1235 s^3 t^2+1115 s^2 t^3\nonumber\\
&+&573 s t^4+92 t^5)+256 m_c^{10} (686 s^6+3499 s^5 t
+6337 s^4 t^2+6952 s^3 t^3+5057 s^2 t^4\nonumber\\
&+&2179 s t^5+446 t^6)-64 m_c^8 (1420 s^7+7165 s^6 t
+15279 s^5 t^2+20080 s^4 t^3+18020 s^3 t^4\nonumber\\
&+&11059 s^2 t^5+4205 s t^6+860 t^7)+16 m_c^6 (1485 s^8
+7888 s^7 t+19416 s^6 t^2+30007 s^5 t^3\nonumber\\
&+&32040 s^4 t^4+24567 s^3 t^5+13136 s^2 t^6+4428 s t^7+845 t^8)
-4 m_c^4 (780 s^9+4522 s^8 t\nonumber\\
&+&12768 s^7 t^2+22889 s^6 t^3+28781 s^5 t^4+26401 s^4 t^5
+17649 s^3 t^6+8348 s^2 t^7+2482 s t^8\nonumber\\
&+&420 t^9)+m_c^2 \left(s^2+s t+t^2\right)^2 (164 s^6
+792 s^5 t+1609 s^4 t^2+1762 s^3 t^3+1289 s^2 t^4\nonumber\\
&+&472 s t^5+84 t^6)-11 s t (s+t) \left(s^2+s t+t^2\right)^4\Big{]}
\end{eqnarray}
\begin{eqnarray}
h_2&=&4 m_c^2 \left( t-4 m_c^2\right)^2 \Big{[}8192 m_c^{14}
(s+t)^2 (19 s+39 t)-2048 m_c^{12} (161 s^4+696 s^3 t\nonumber\\
&+&1206 s^2 t^2+928 s t^3+257 t^4)+512 m_c^{10}
(473 s^5+2320 s^4 t+5423 s^3 t^2+6065 s^2 t^3\nonumber\\
&+&3340 s t^4+723 t^5)-128 m_c^8 (623 s^6+3752 s^5 t
+11313 s^4 t^2+16704 s^3 t^3+13939 s^2 t^4\nonumber\\
&+&6152 s t^5+1149 t^6)+32 m_c^6 (376 s^7+3161 s^6 t
+11876 s^5 t^2+22019 s^4 t^3+24449 s^3 t^4\nonumber\\
&+&16498 s^2 t^5+6351 s t^6+1086 t^7)-8 m_c^4 (84 s^8
+1328 s^7 t+6071 s^6 t^2+13796 s^5 t^3\nonumber\\
&+&18908 s^4 t^4+17172 s^3 t^5+10043 s^2 t^6+3536 s t^7
+566 t^8)+2 m_c^2 t (218 s^8+1208 s^7 t\nonumber\\
&+&3361 s^6 t^2+5431 s^5 t^3+6167 s^4 t^4+4797 s^3 t^5
+2558 s^2 t^6+832 s t^7+124 t^8)\nonumber\\
&-&s t^2 \left(s^2+s t+t^2\right)^2 \left(7 s^3+14 s^2 t
+9 s t^2+2 t^3\right)\Big{]}
\end{eqnarray}
\begin{eqnarray}
h_3&=&-4 m_c^2 (s+t)^2 \Big{[}262144 m_c^{16} \left(s^2+t^2\right)
-8192 m_c^{14} \left(9 s^3-19 s^2 t+21 s t^2+9 t^3\right)\nonumber\\
&-&2048 m_c^{12} \left(157 s^4+400 s^3 t+218 s^2 t^2+120 s t^3
+137 t^4\right)+512 m_c^{10} (683 s^5+1914 s^4 t\nonumber\\
&+&1569 s^3 t^2+849 s^2 t^3+994 s t^4+563 t^5)-128 m_c^8
(1321 s^6+4100 s^5 t+4181 s^4 t^2\nonumber\\
&+&2372 s^3 t^3+2001 s^2 t^4+2340 s t^5+1021 t^6)+32 m_c^6
(1382 s^7+4555 s^6 t+5490 s^5 t^2\nonumber\\
&+&3491 s^4 t^3+2011 s^3 t^4+2610 s^2 t^5+2595 s t^6+1022 t^7)
-8m_c^4 (750 s^8+2584 s^7 t\nonumber\\
&+&3563 s^6 t^2+2580 s^5 t^3+1242 s^4 t^4+1100 s^3 t^5+1703 s^2 t^6
+1424 s t^7+550 t^8)\nonumber\\
&+&2m_c^2 (164 s^9+600 s^8 t+926 s^7 t^2+753 s^6 t^3
+383 s^5 t^4+183 s^4 t^5+313 s^3 t^6+446 s^2 t^7\nonumber\\
&+&320 s t^8+124 t^9)-s t \left(2 s^2-3 s t+2 t^2\right)
\left(s^3+2 s^2 t+2 s t^2+t^3\right)^2\Big{]}
\end{eqnarray}
\begin{eqnarray}
h_4&=&4 m_c^2 \left(4 m_c^2-s\right) \Big{[}8192 m_c^{14}
\left(13 s^4+76 s^3 t+42 s^2 t^2-52 s t^3-31 t^4\right)\nonumber\\
&-&4096 m_c^{12} \left(92 s^5+435 s^4 t+486 s^3 t^2+13 s^2 t^3
-322 s t^4-128 t^5\right)+512 m_c^{10} (729 s^6\nonumber\\
&+&3578 s^5 t+6026 s^4 t^2+3382 s^3 t^3-1629 s^2 t^4-2648 s t^5
-846 t^6)-256 m_c^8 (664 s^7\nonumber\\
&+&3613 s^6 t+7719 s^5 t^2+7378 s^4 t^3+1751 s^3 t^4
-2993 s^2 t^5-2634 s t^6-722 t^7)\nonumber\\
&+&32 m_c^6 (1324 s^8+7870 s^7 t+19424 s^6 t^2
+24728 s^5 t^3+14981 s^4 t^4-1074 s^3 t^5\nonumber\\
&-&8562 s^2 t^6-5580 s t^7-1351 t^8)-16 m_c^4 (359 s^9
+2242 s^8 t+6074 s^7 t^2+9229 s^6 t^3\nonumber\\
&+&7941 s^5 t^4+2897 s^4 t^5-1764 s^3 t^6-2790 s^2 t^7
-1522 s t^8-330 t^9)+2 m_c^2 (164 s^{10}\nonumber\\
&+&1060 s^9 t+3088 s^8 t^2+5238 s^7 t^3+5563 s^6 t^4+3338 s^5 t^5
+368 s^4 t^6-1476 s^3 t^7-1411 s^2 t^8\nonumber\\
&-&680 s t^9-132 t^{10})-s t \left(s^2+s t+t^2\right)^2
\left(2 s^5+7 s^4 t+8 s^3 t^2-s^2 t^3-6 s t^4-2 t^5\right)\Big{]}
\end{eqnarray}
\end{subequations}

$g+g\to c\bar{c}({}^3\!P_2^{[1]})+g$:
\begin{subequations}
\begin{eqnarray}
a^{gg}({}^3\!P_2^{[1]})
=-\frac{16 \pi ^3 \alpha_s^3}{135 m_c^3 s t \left(s-4 m_c^2\right)^4
\left(t-4 m_c^2\right)^4 (s+t)^4 \left(-4 m_c^2+s+t\right)}
\end{eqnarray}
\begin{eqnarray}
g_1&=&-18874368 m_c^{20} (s+t)^4+9437184 m_c^{18} (s+t)^3
\left(3 s^2+5 s t+3 t^2\right)
\nonumber\\
&-&16384 m_c^{16} (s+t)^2 \left(1224 s^4+3897 s^3 t
+5297 s^2 t^2+3897 s t^3+1224 t^4\right)\nonumber\\
&+&4096 m_c^{14} (2160 s^7+11913 s^6 t+29647 s^5 t^2
+44652 s^4 t^3+44652 s^3 t^4+29647 s^2 t^5\nonumber\\
&+&11913 s t^6+2160 t^7)-1024 m_c^{12} (2592 s^8
+14994 s^7 t+39656 s^6 t^2+65365 s^5 t^3\nonumber\\
&+&76238 s^4 t^4+65365 s^3 t^5+39656 s^2 t^6+14994 s t^7+2592 t^8)
+256 m_c^{10} (2160 s^9\nonumber\\
&+&13374 s^8 t+37078 s^7 t^2+63603 s^6 t^3+79713 s^5 t^4
+79713 s^4 t^5+63603 s^3 t^6+37078 s^2 t^7\nonumber\\
&+&13374 s t^8+2160 t^9)-64 m_c^8 (1224 s^{10}
+8445 s^9 t+24967 s^8 t^2+43880 s^7 t^3+55214 s^6 t^4\nonumber\\
&+&58192 s^5 t^5+55214 s^4 t^6+43880 s^3 t^7+24967 s^2 t^8
+8445 s t^9+1224 t^{10})+16 m_c^6 (432 s^{11}\nonumber\\
&+&3501 s^{10} t+11595 s^9 t^2+21568 s^8 t^3+26848 s^7 t^4
+27020 s^6 t^5+27020 s^5 t^6+26848 s^4 t^7\nonumber\\
&+&21568 s^3 t^8+11595 s^2 t^9+3501 s t^{10}+432 t^{11})-4 m_c^4
(72 s^{12}+792 s^{11} t+3218 s^{10} t^2\nonumber\\
&+&6815 s^9 t^3+8979 s^8 t^4+8574 s^7 t^5+7840 s^6 t^6
+8574 s^5 t^7+8979 s^4 t^8+6815 s^3 t^9\nonumber\\
&+&3218 s^2 t^{10}+792 s t^{11}+72 t^{12})+m_c^2 s t \left(s^2+s t
+t^2\right)^2(60 s^7+264 s^6 t+341 s^5 t^2+71 s^4 t^3\nonumber\\
&+&71 s^3 t^4+341 s^2 t^5+264 s t^6+60 t^7)
-s^2 t^2 (s+t)^2 \left(s^2+s t+t^2\right)^4
\end{eqnarray}
\begin{eqnarray}
g_2&=&-12 m_c^2 \left(4 m_c^2-t\right) \Big{[}8192 m_c^{14}
(s+t)^2 \left(12 s^2+47 s t+24 t^2\right)-2048 m_c^{12} (56 s^5\nonumber\\
&+&345 s^4 t+722 s^3 t^2+717 s^2 t^3+352 s t^4+68 t^5)
+512 m_c^{10} (112 s^6+741 s^5 t+1745 s^4 t^2\nonumber\\
&+&2129 s^3 t^3+1501 s^2 t^4+552 s t^5+72 t^6)-128 m_c^8
(128 s^7+919 s^6 t+2368 s^5 t^2\nonumber\\
&+&3306 s^4 t^3+2900 s^3 t^4+1591 s^2 t^5+480 s t^6+48 t^7)
+32 m_c^6 (92 s^8+704 s^7 t+1933 s^6 t^2\nonumber\\
&+&2902 s^5 t^3+2844 s^4 t^4+1945 s^3 t^5+905 s^2 t^6+259 s t^7
+40 t^8)-8 m_c^4 (40 s^9+324 s^8 t\nonumber\\
&+&952 s^7 t^2+1461 s^6 t^3+1416 s^5 t^4+983 s^4 t^5+558 s^3 t^6
+298 s^2 t^7+132 s t^8+36 t^9)\nonumber\\
&+&2 m_c^2 (8 s^{10}+80 s^9 t+274 s^8 t^2+432 s^7 t^3
+355 s^6 t^4+133 s^5 t^5+5 s^4 t^6+71 s^3 t^7+132 s^2 t^8\nonumber\\
&+&94 s t^9+24 t^{10})-t \left(s^2+s t+t^2\right)^2 \left(4 s^6
+12 s^5 t+s^4 t^2-18 s^3 t^3-s^2 t^4+10 s t^5+4 t^6\right)\Big{]}\nonumber\\
\end{eqnarray}
\begin{eqnarray}
g_3&=&-12 m_c^2 (s+t) \Big{[}1572864 m_c^{16} (s+t)^3
-294912 m_c^{14} (s+t)^2 \left(7 s^2+10 s t+7 t^2\right)\nonumber\\
&+&2048 m_c^{12} \left(588 s^5+2257 s^4 t+3917 s^3 t^2+3917 s^2 t^3
+2257 s t^4+588 t^5\right)\nonumber\\
&-&512 m_c^{10} \left(824 s^6+3435 s^5 t+6696 s^4 t^2+8202 s^3 t^3
+6696 s^2 t^4+3435 s t^5+824 t^6\right)\nonumber\\
&+&128 m_c^8 (784 s^7+3461 s^6 t+7141 s^5 t^2+9650 s^4 t^3
+9650 s^3 t^4+7141 s^2 t^5+3461 s t^6\nonumber\\
&+&784 t^7)-32 m_c^6 (524 s^8+2459 s^7 t+5234 s^6 t^2
+7181 s^5 t^3+7796 s^4 t^4+7181 s^3 t^5\nonumber\\
&+&5234 s^2 t^6+2459 s t^7+524 t^8)+8 m_c^4 (236 s^9+1218 s^8 t
+2722 s^7 t^2+3705 s^6 t^3\nonumber\\
&+&3905 s^5 t^4+3905 s^4 t^5+3705 s^3 t^6+2722 s^2 t^7+1218 s t^8
+236 t^9)-2 m_c^2 (s+t)^2 (64 s^8\nonumber\\
&+&246 s^7 t+356 s^6 t^2+337 s^5 t^3+274 s^4 t^4+337 s^3 t^5
+356 s^2 t^6+246 s t^7+64 t^8)\nonumber\\
&+&\left(s^2+s t+t^2\right)^2 \left(4 s^7+18 s^6 t+23 s^5 t^2+3 s^4 t^3
+3 s^3 t^4+23 s^2 t^5+18 s t^6+4 t^7\right)\Big{]}
\end{eqnarray}
\begin{eqnarray}
g_4&=&-12 m_c^2 \Big{[}196608 m_c^{16} (s+t)^3 (3 s+5 t)
-73728 m_c^{14} (s+t)^2 (12 s^3+33 s^2 t+35 s t^2\nonumber\\
&+&16 t^3)+2048 m_c^{12} (280 s^6+1369 s^5 t+3023 s^4 t^2
+3917 s^3 t^3+3151 s^2 t^4+1476 s t^5\nonumber\\
&+&308 t^6)-512 m_c^{10} (416 s^7+2129 s^6 t+5074 s^5 t^2
+7460 s^4 t^3+7438 s^3 t^4+5057 s^2 t^5\nonumber\\
&+&2130 s t^6+408 t^7)+128 m_c^8 (394 s^8+2169 s^7 t
+5479 s^6 t^2+8608 s^5 t^3+9650 s^4 t^4\nonumber\\
&+&8183 s^3 t^5+5123 s^2 t^6+2076 s t^7+390 t^8)-32 m_c^6
(244 s^9+1504 s^8 t+4073 s^7 t^2\nonumber\\
&+&6653 s^6 t^3+7734 s^5 t^4+7243 s^4 t^5+5762 s^3 t^6+3620 s^2 t^7
+1479 s t^8+280 t^9)\nonumber\\
&+&8 m_c^4 (92 s^{10}+674 s^9 t+2054 s^8 t^2+3591 s^7 t^3
+4219 s^6 t^4+3905 s^5 t^5+3391 s^4 t^6\nonumber\\
&+&2836 s^3 t^7+1886 s^2 t^8+780 s t^9+144 t^{10})-2 m_c^2
(16 s^{11}+162 s^{10} t+606 s^9 t^2+1224 s^8 t^3\nonumber\\
&+&1555 s^7 t^4+1400 s^6 t^5+1126 s^5 t^6+1044 s^4 t^7+983 s^3 t^8
+680 s^2 t^9+276 s t^{10}+48 t^{11})\nonumber\\
&+&t \left(s^2+s t+t^2\right)^2 \left(6 s^7+23 s^6 t+28 s^5 t^2
+3 s^4 t^3-2 s^3 t^4+18 s^2 t^5+16 s t^6+4 t^7\right)\Big{]}
\end{eqnarray}
\begin{eqnarray}
b^{gg}({}^3\!P_2^{[1]})
=\frac{8 \pi ^3 \alpha_s^3}{675 m_c^5 s^2 t^2 \left(4 m_c^2-s\right)^5
\left(4 m_c^2-t\right)^5 (s+t)^5 \left(-4 m_c^2+s+t\right)^2}
\end{eqnarray}
\begin{eqnarray}
h_1&=&-s t \left(-4 m_c^2+s+t\right) \Big{[}301989888 (s+t)^3
\left(15 s^2+38 t s+15 t^2\right) m_c^{24}\nonumber\\
&-&8388608 (s+t)^2 \left(945 s^4+3943 t s^3+5985 t^2 s^2+3763 t^3 s
+945 t^4\right) m_c^{22}\nonumber\\
&+&262144 (27624 s^7+179241 t s^6+503190 t^2 s^5+788233 t^3 s^4
+755573 t^4 s^3+450370 t^5 s^2\nonumber\\
&+&156201 t^6 s+24744 t^7) m_c^{20}-65536 (65928 s^8
+438138 t s^7+1296615 t^2 s^6\nonumber\\
&+&2241746 t^3 s^5+2524170 t^4 s^4+1935826 t^5 s^3+1003615 t^6 s^2
+325218 t^7 s+50088 t^8) m_c^{18}\nonumber\\
&+&16384 (108144 s^9+742491 t s^8+2285700 t^2 s^7
+4194310 t^3 s^6+5180803 t^4 s^5\nonumber\\
&+&4615783 t^5 s^4+3046670 t^6 s^3+1457000 t^7 s^2
+456891 t^8 s+69264 t^9) m_c^{16}\nonumber\\
&-&4096 (124080 s^{10}+900416 t s^9+2906497 t^2 s^8
+5577856 t^3 s^7+7244723 t^4 s^6\nonumber\\
&+&6963392 t^5 s^5+5278403 t^6 s^4+3209416 t^7 s^3+1491977 t^8 s^2
+460856 t^9 s+67920 t^{10}) m_c^{14}\nonumber\\
&+&1024 (99240 s^{11}+779395 t s^{10}+2690450 t^2 s^9
+5443508 t^3 s^8+7327604 t^4 s^7\nonumber\\
&+&7226103 t^5 s^6+5774383 t^6 s^5+4032804 t^7 s^4+2433368 t^8 s^3
+1137870 t^9 s^2+343315 t^{10} s\nonumber\\
&+&47400 t^{11}) m_c^{12}-256 (53064 s^{12}+463250 t s^{11}
+1757329 t^2 s^{10}+3842678 t^3 s^9\nonumber\\
&+&5442560 t^4 s^8+5431020 t^5 s^7+4269078 t^6 s^6+3095980 t^7 s^5
+2225920 t^8 s^4+1409878 t^9 s^3\nonumber\\
&+&655689 t^{10} s^2+185450 t^{11} s+22824 t^{12}) m_c^{10}+64
(16968 s^{13}+173309 t s^{12}+751960 t^2 s^{11}\nonumber\\
&+&1842678 t^3 s^{10}+2850960 t^4 s^9+2968913 t^5 s^8
+2256004 t^6 s^7+1525084 t^7 s^6\nonumber\\
&+&1187673 t^8 s^5+968900 t^9 s^4+624838 t^{10} s^3+270020 t^{11} s^2
+66749 t^{12} s+6888 t^{13}) m_c^8\nonumber\\
&-&16 (2424 s^{14}+34060 t s^{13}+183551 t^2 s^{12}+532924 t^3 s^{11}
+949514 t^4 s^{10}+1102104 t^5 s^9\nonumber\\
&+&872595 t^6 s^8+540792 t^7 s^7+397555 t^8 s^6+391504 t^9 s^5
+330714 t^{10} s^4+190964 t^{11} s^3\nonumber\\
&+&68951 t^{12} s^2+13300 t^{13} s+984 t^{14}) m_c^6+4 s t
(2116 s^{13}+19052 t s^{12}+74123 t^2 s^{11}\nonumber\\
&+&165228 t^3 s^{10}+235243 t^4 s^9+227226 t^5 s^8
+164836 t^6 s^7+119456 t^7 s^6+113946 t^8 s^5\nonumber\\
&+&107443 t^9 s^4+73328 t^{10} s^3+32643 t^{11} s^2+8372 t^{12} s
+916 t^{13}) m_c^4-s^2 t^2 (s^3+2 t s^2\nonumber\\
&+&2 t^2 s+t^3)^2 \left(360 s^6+1356 t s^5+1421 t^2 s^4+494 t^3 s^3
+1101 t^4 s^2+1036 t^5 s+280 t^6\right) m_c^2\nonumber\\
&+&7 s^3 t^3 (s+t)^3 \left(s^2+t s+t^2\right)^4\Big{]}
\end{eqnarray}
\begin{eqnarray}
h_2&=&-12 m_c^2 \left(4 m_c^2-t\right) \Big{[}16777216 s t (s+t)^2
\left(8 s^2-9 t s+2 t^2\right) m_c^{22}+524288 (32 s^7\nonumber\\
&-&636 t s^6-2381 t^2 s^5-4579 t^3 s^4-5103 t^4 s^3-2877 t^5 s^2
-608 t^6 s+32 t^7) m_c^{20}\nonumber\\
&-&131072 (224 s^8-2256 t s^7-13229 t^2 s^6-34841 t^3 s^5
-50353 t^4 s^4-39495 t^5 s^3-16038 t^6 s^2\nonumber\\
&-&2292 t^7 s+256 t^8) m_c^{18}+32768 (672 s^9-4044 t s^8
-34626 t^2 s^7-111888 t^3 s^6\nonumber\\
&-&195319 t^4 s^5-196987 t^5 s^4-116723 t^6 s^3-35861 t^7 s^2
-2592 t^8 s+896 t^9) m_c^{16}\nonumber\\
&-&8192 (1120 s^{10}-4264 t s^9-54962 t^2 s^8-207478 t^3 s^7
-416241 t^4 s^6-501995 t^5 s^5\nonumber\\
&-&383465 t^6 s^4-180583 t^7 s^3-41448 t^8 s^2+1468 t^9 s
+1792 t^{10}) m_c^{14}+2048 (1120 s^{11}\nonumber\\
&-&3164 t s^{10}-59449 t^2 s^9-249681 t^3 s^8-549542 t^4 s^7
-743186 t^5 s^6-667344 t^6 s^5\nonumber\\
&-&410272 t^7 s^4-160085 t^8 s^3-25309 t^9 s^2+6312 t^{10} s
+2240 t^{11}) m_c^{12}-512 (672 s^{12}\nonumber\\
&-&2176 t s^{11}-46457 t^2 s^{10}-206373 t^3 s^9-475554 t^4 s^8
-672814 t^5 s^7-639440 t^6 s^6\nonumber\\
&-&444400 t^7 s^5-235223 t^8 s^4-84109 t^9 s^3-8818 t^{10} s^2
+6012 t^{11} s+1792 t^{12}) m_c^{10}\nonumber\\
&+&128 (224 s^{13}-1396 t s^{12}-26208 t^2 s^{11}
-119918 t^3 s^{10}-279940 t^4 s^9-383196 t^5 s^8\nonumber\\
&-&319618 t^6 s^7-174546 t^7 s^6-88697 t^8 s^5-63753 t^9 s^4
-36485 t^{10} s^3-7707 t^{11} s^2\nonumber\\
&+&1808 t^{12} s+896 t^{13}) m_c^8-32 (32 s^{14}-584 t s^{13}
-9872 t^2 s^{12}-47988 t^3 s^{11}-115308 t^4 s^{10}\nonumber\\
&-&146248 t^5 s^9-68062 t^6 s^8+62478 t^7 s^7
+108433 t^8 s^6+47003 t^9 s^5-20127 t^{10} s^4\nonumber\\
&-&29125 t^{11} s^3-11012 t^{12} s^2-972 t^{13} s+256 t^{14}) m_c^6-8 t
(104 s^{14}+2056 t s^{13}+11846 t^2 s^{12}\nonumber\\
&+&32068 t^3 s^{11}+41516 t^4 s^{10}+3659 t^5 s^9-73197 t^6 s^8
-117854 t^7 s^7-85086 t^8 s^6\nonumber\\
&-&15497 t^9 s^5+23451 t^{10} s^4+20944 t^{11} s^3+7342 t^{12} s^2
+952 t^{13} s-32 t^{14}) m_c^4\nonumber\\
&+&2 s t^2 (144 s^{13}+1304 t s^{12}+4612 t^2 s^{11}+7188 t^3 s^{10}
+183 t^4 s^9-19901 t^5 s^8-38638 t^6 s^7\nonumber\\
&-&36862 t^7 s^6-15497 t^8 s^5+4991 t^9 s^4+10688 t^{10} s^3
+6400 t^{11} s^2+1860 t^{12} s+232 t^{13}) m_c^2\nonumber\\
&-&s^2 t^3 \left(s^3+2 t s^2+2 t^2 s+t^3\right)^2 \left(4 s^6+12 t s^5
-83 t^2 s^4-186 t^3 s^3-21 t^4 s^2+74 t^5 s+20 t^6\right)\Big{]}\nonumber\\
\end{eqnarray}
\begin{eqnarray}
h_3&=&12 m_c^2 (s+t) \Big{[}1207959552 s t (s+t)^2 \left(2 s^2
+5 t s+2 t^2\right) m_c^{22}-2097152 s t (2003 s^5\nonumber\\
&+&10179 t s^4+20112 t^2 s^3+19932 t^3 s^2+10089 t^4 s+2093 t^5)
m_c^{20}-131072 (32 s^8\nonumber\\
&-&26276 t s^7-146167 t^2 s^6-337414 t^3 s^5
-428510 t^4 s^4-324974 t^5 s^3-140887 t^6 s^2\nonumber\\
&-&27476 t^7 s+32 t^8) m_c^{18}+32768 (224 s^9
-54588 t s^8-323187 t^2 s^7-823603 t^3 s^6\nonumber\\
&-&1219834 t^4 s^5-1174194 t^5 s^4-747863 t^6 s^3
-293687 t^7 s^2-54228 t^8 s+224 t^9) m_c^{16}\nonumber\\
&-&8192 (672 s^{10}-80108 t s^9-497784 t^2 s^8-1348997 t^3 s^7
-2185606 t^4 s^6-2420290 t^5 s^5\nonumber\\
&-&1940166 t^6 s^4-1111197 t^7 s^3-411864 t^8 s^2-72268 t^9 s
+672 t^{10}) m_c^{14}+4096 (560 s^{11}\nonumber\\
&-&43042 t s^{10}-281992 t^2 s^9-800424 t^3 s^8-1361965 t^4 s^7
-1614273 t^5 s^6-1466973 t^6 s^5\nonumber\\
&-&1057525 t^7 s^4-577964 t^8 s^3-207612 t^9 s^2-34462 t^{10} s
+560 t^{11}) m_c^{12}-512 (1120 s^{12}\nonumber\\
&-&67652 t s^{11}-474251 t^2 s^{10}-1417688 t^3 s^9-2502261 t^4 s^8
-3032698 t^5 s^7-2861316 t^6 s^6\nonumber\\
&-&2322578 t^7 s^5-1630861 t^8 s^4-889688 t^9 s^3-311491 t^{10} s^2
-48212 t^{11} s+1120 t^{12}) m_c^{10}\nonumber\\
&+&128 (672 s^{13}-38036 t s^{12}-288627 t^2 s^{11}-922087 t^3 s^{10}
-1708003 t^4 s^9-2099899 t^5 s^8\nonumber\\
&-&1933612 t^6 s^7-1574572 t^7 s^6-1275799 t^8 s^5-940063 t^9 s^4
-517827 t^{10} s^3-173847 t^{11} s^2\nonumber\\
&-&24636 t^{12} s+672 t^{13}) m_c^8-32 (224 s^{14}
-14628 t s^{13}-120242 t^2 s^{12}-415269 t^3 s^{11}\nonumber\\
&-&821945 t^4 s^{10}-1045419 t^5 s^9-925223 t^6 s^8-670140 t^7 s^7
-542503 t^8 s^6-510619 t^9 s^5\nonumber\\
&-&406585 t^{10} s^4-220149 t^{11} s^3-69042 t^{12} s^2-8868 t^{13} s
+224 t^{14}) m_c^6+8 (32 s^{15}\nonumber\\
&-&3488 t s^{14}-30854 t^2 s^{13}-115762 t^3 s^{12}-247199 t^4 s^{11}
-330131 t^5 s^{10}-280925 t^6 s^9\nonumber\\
&-&153705 t^7 s^8-83045 t^8 s^7-107385 t^9 s^6-142151 t^{10} s^5
-119459 t^{11} s^4-61242 t^{12} s^3\nonumber\\
&-&17534 t^{13} s^2-2048 t^{14} s+32 t^{15}) m_c^4+2 s t (s+t)^2
(392 s^{12}+2996 t s^{11}+9224 t^2 s^{10}\nonumber\\
&+&14502 t^3 s^9+10803 t^4 s^8-1352 t^5 s^7-10146 t^6 s^6-7032 t^7 s^5
+2203 t^8 s^4+6782 t^9 s^3\nonumber\\
&+&5064 t^{10} s^2+1756 t^{11} s+232 t^{12}) m_c^2-s^2 t^2 (s+t)^3
\left(s^2+t s+t^2\right)^2(20 s^6+46 t s^5\nonumber\\
&-&91 t^2 s^4-238 t^3 s^3-91 t^4 s^2+46 t^5 s+20 t^6)\Big{]}
\end{eqnarray}
\begin{eqnarray}
h_4&=&-12 m_c^2 \Big{[}201326592 s (s-t) t (s+t)^3 m_c^{24}
-4194304 s t (s+t)^2 (291 s^3+899 t s^2\nonumber\\
&+&997 t^2 s+285 t^3) m_c^{22}+524288 (32 s^8+3928 t s^7
+22769 t^2 s^6+56744 t^3 s^5+76938 t^4 s^4\nonumber\\
&+&59740 t^5 s^3+25149 t^6 s^2+4444 t^7 s-32 t^8)
m_c^{20}-131072 (224 s^9+13750 t s^8\nonumber\\
&+&84023 t^2 s^7+231650 t^3 s^6+369902 t^4 s^5+371442 t^5 s^4
+235061 t^6 s^3+86390 t^7 s^2\nonumber\\
&+&13894 t^8 s-256 t^9) m_c^{18}+32768 (672 s^{10}
+30316 t s^9+191386 t^2 s^8+564744 t^3 s^7\nonumber\\
&+&1001907 t^4 s^6+1182124 t^5 s^5+960430 t^6 s^4
+525736 t^7 s^3+174749 t^8 s^2+25168 t^9 s\nonumber\\
&-&896 t^{10}) m_c^{16}-8192 (1120 s^{11}+45670 t s^{10}
+298834 t^2 s^9+931048 t^3 s^8+1770223 t^4 s^7\nonumber\\
&+&2303172 t^5 s^6+2179344 t^6 s^5+1526640 t^7 s^4
+760913 t^8 s^3+236598 t^9 s^2+30486 t^{10} s\nonumber\\
&-&1792 t^{11}) m_c^{14}+2048 (1120 s^{12}+48192 t s^{11}
+333351 t^2 s^{10}+1101380 t^3 s^9+2213799 t^4 s^8\nonumber\\
&+&3049824 t^5 s^7+3122706 t^6 s^6+2517072 t^7 s^5
+1614829 t^8 s^4+772192 t^9 s^3+231987 t^{10} s^2\nonumber\\
&+&27212 t^{11} s-2240 t^{12}) m_c^{12}-512 (672 s^{13}
+35098 t s^{12}+264391 t^2 s^{11}+944150 t^3 s^{10}\nonumber\\
&+&2026693 t^4 s^9+2927376 t^5 s^8+3093530 t^6 s^7
+2615244 t^7 s^6+1916873 t^8 s^5+1212706 t^9 s^4\nonumber\\
&+&586319 t^{10} s^3+174642 t^{11} s^2+19394 t^{12} s-1792 t^{13})
m_c^{10}+128 (224 s^{14}+16740 t s^{13}\nonumber\\
&+&142436 t^2 s^{12}+565020 t^3 s^{11}+1329468 t^4 s^{10}
+2056868 t^5 s^9+2236868 t^6 s^8+1858992 t^7 s^7\nonumber\\
&+&1367893 t^8 s^6+1012724 t^9 s^5+691232 t^{10} s^4+349104 t^{11} s^3
+103807 t^{12} s^2+11424 t^{13} s\nonumber\\
&-&896 t^{14}) m_c^8-32 (32 s^{15}+4676 t s^{14}
+47458 t^2 s^{13}+216720 t^3 s^{12}+577256 t^4 s^{11}\nonumber\\
&+&990974 t^5 s^{10}+1149214 t^6 s^9+936240 t^7 s^8
+606773 t^8 s^7+438198 t^9 s^6+396540 t^{10} s^5\nonumber\\
&+&310428 t^{11} s^4+161311 t^{12} s^3+47012 t^{13} s^2+5248 t^{14} s
-256 t^{15}) m_c^6+8 t (576 s^{15}\nonumber\\
&+&8074 t s^{14}+45396 t^2 s^{13}+142830 t^3 s^{12}+281476 t^4 s^{11}
+360889 t^5 s^{10}+296852 t^6 s^9\nonumber\\
&+&148525 t^7 s^8+60588 t^8 s^7+78887 t^9 s^6+117204 t^{10} s^5
+101911 t^{11} s^4+51660 t^{12} s^3\nonumber\\
&+&14268 t^{13} s^2+1600 t^{14} s-32 t^{15}) m_c^4-2 s t^2 (s+t)^2
(396 s^{12}+2764 t s^{11}+8634 t^2 s^{10}\nonumber\\
&+&13982 t^3 s^9+10851 t^4 s^8-1158 t^5 s^7-10180 t^6 s^6
-7280 t^7 s^5+2283 t^8 s^4+6972 t^9 s^3\nonumber\\
&+&5136 t^{10} s^2+1712 t^{11} s+232 t^{12}) m_c^2
+s^2 t^3 (s+t)^3 \left(s^2+t s+t^2\right)^2\nonumber\\
&&\left(6 s^6+17 t s^5-105 t^2 s^4-224 t^3 s^3-62 t^4 s^2
+60 t^5 s+20 t^6\right)\Big{]}
\end{eqnarray}
\end{subequations}

$g+g\to c\bar{c}({}^3\!S_1^{[8]})+g$:
\begin{subequations}
\begin{eqnarray}
b^{gg}({}^3\!S_1^{[8]})
=-\frac{\pi ^3 \alpha_s^3}{108 m_c^5 s t \left(4 m_c^2-s\right)^3
\left(4 m_c^2-t\right)^3 (s+t)^3 \left(-4 m_c^2+s+t\right)}
\end{eqnarray}
\begin{eqnarray}
h_1&=&-s t \left(-4 m_c^2+s+t\right) \Big{[}622592 m_c^{14}
\left(3 s^2+2 s t+3 t^2\right)-4096 m_c^{12} (203 s^3-12 s^2 t\nonumber\\
&+&216 s t^2+203 t^3)-1024 m_c^{10} \left(420 s^4+2357 s^3 t
+2358 s^2 t^2+1187 s t^3+258 t^4\right)\nonumber\\
&+&256 m_c^8 \left(2073 s^5+8653 s^4 t+12881 s^3 t^2+10805 s^2 t^3
+5215 s t^4+1263 t^5\right)\nonumber\\
&-&64 m_c^6 \left(3265 s^6+14055 s^5 t+26073 s^4 t^2+28978 s^3 t^3
+20175 s^2 t^4+8319 s t^5+1807 t^6\right)\nonumber\\
&+&48 m_c^4 (774 s^7+3736 s^6 t+8400 s^5 t^2+11507 s^4 t^3
+10373 s^3 t^4+6186 s^2 t^5+2170 s t^6\nonumber\\
&+&396 t^7)-108 m_c^2 (23 s^8+142 s^7 t+392 s^6 t^2
+653 s^5 t^3+742 s^4 t^4+569 s^3 t^5+296 s^2 t^6\nonumber\\
&+&88 s t^7+11 t^8)+297 s t (s+t) \left(s^2+s t+t^2\right)^3\Big{]}
\end{eqnarray}
\begin{eqnarray}
h_2&=&8 m_c^2 t \Big{[}4980736 m_c^{16} s-16384 m_c^{14}
\left(214 s^2+13 s t+27 t^2\right)+4096 m_c^{12} (125 s^3\nonumber\\
&-&1442 s^2 t-799 s t^2+216 t^3)+1024 m_c^{10} (1299 s^4
+7096 s^3 t+7922 s^2 t^2+1845 s t^3\nonumber\\
&-&756 t^4)-256 m_c^8 \left(4029 s^5+17654 s^4 t+25019 s^3 t^2
+14306 s^2 t^3+1152 s t^4-1512 t^5\right)\nonumber\\
&+&64 m_c^6 \left(5497 s^6+24150 s^5 t+40980 s^4 t^2+33774 s^3 t^3
+11507 s^2 t^4-1511 s t^5-1863 t^6\right)\nonumber\\
&-&16 m_c^4 (3672 s^7+17571 s^6 t+35040 s^5 t^2+36622 s^4 t^3
+19839 s^3 t^4+2237 s^2 t^5-3125 s t^6\nonumber\\
&-&1404 t^7)+4 m_c^2 (918 s^8+5400 s^7 t+13063 s^6 t^2
+16698 s^5 t^3+11639 s^4 t^4+2810 s^3 t^5\nonumber\\
&-&2283 s^2 t^6-2057 s t^7-594 t^8)+27 t (-10 s^8-40 s^7 t
-63 s^6 t^2-49 s^5 t^3+2 s^4 t^4+39 s^3 t^5\nonumber\\
&+&39 s^2 t^6+18 s t^7+4 t^8)\Big{]}
\end{eqnarray}
\begin{eqnarray}
h_3&=&8 m_c^2 \left(4 m_c^2-s-t\right) \Big{[}1024 m_c^{10}
\left(27 s^4+260 s^3 t+162 s^2 t^2+260 s t^3+27 t^4\right)\nonumber\\
&-&256 m_c^8 \left(162 s^5+617 s^4 t+939 s^3 t^2+1395 s^2 t^3+617 s t^4
+162 t^5\right)+64 m_c^6 (405 s^6\nonumber\\
&+&1238 s^5 t+1841 s^4 t^2+3670 s^3 t^3+2717 s^2 t^4+1172 s t^5+405 t^6)
-32 m_c^4 (270 s^7\nonumber\\
&+&642 s^6 t+1190 s^5 t^2+2225 s^4 t^3+2387 s^3 t^4+1757 s^2 t^5+723 s t^6
+270 t^7)+4 m_c^2 (378 s^8\nonumber\\
&+&1193 s^7 t+2369 s^6 t^2+3793 s^5 t^3+4964 s^4 t^4+4117 s^3 t^5
+2855 s^2 t^6+1193 s t^7+378 t^8)\nonumber\\
&-&27 \left(4 s^9+18 s^8 t+39 s^7 t^2+66 s^6 t^3+83 s^5 t^4+83 s^4 t^5
+66 s^3 t^6+39 s^2 t^7+18 s t^8+4 t^9\right)\Big{]}\nonumber\\
\end{eqnarray}
\begin{eqnarray}
h_4&=&8 m_c^2 \Big{[}221184 m_c^{14} (s-t) (s+t)^2-2048 m_c^{12}
(189 s^4-110 s^3 t-162 s^2 t^2-410 s t^3\nonumber\\
&-&243 t^4)+512 m_c^{10} \left(567 s^5+566 s^4 t-495 s^3 t^2-2227 s^2 t^3
-2374 s t^4-945 t^5\right)\nonumber\\
&-&128 m_c^8 \left(945 s^6+2744 s^5 t+2585 s^4 t^2-4606 s^3 t^3
-9379 s^2 t^4-6778 s t^5-2079 t^6\right)\nonumber\\
&+&32 m_c^6 (918 s^7+4957 s^6 t+8354 s^5 t^2-709 s^4 t^3
-15709 s^3 t^4-19272 s^2 t^5-10811 s t^6\nonumber\\
&-&2808 t^7)-8 m_c^4 (486 s^8+3942 s^7 t+9139 s^6 t^2
+4902 s^5 t^3-11758 s^4 t^4-23880 s^3 t^5\nonumber\\
&-&21205 s^2 t^6-9976 s t^7-2322 t^8)+2 m_c^2 (108 s^9
+918 s^8 t+2906 s^7 t^2+2147 s^6 t^3\nonumber\\
&-&5453 s^5 t^4-14923 s^4 t^5-17387 s^3 t^6-12136 s^2 t^7-5032 s t^8
-1080 t^9)+27 t (2 s^9+4 s^8 t\nonumber\\
&+&13 s^7 t^2+42 s^6 t^3+83 s^5 t^4+107 s^4 t^5+92 s^3
   t^6+53 s^2 t^7+20 s t^8+4 t^9)\Big{]}
\end{eqnarray}
\end{subequations}

$g+g\to c\bar{c}({}^3\!P_J^{[8]})+g$:
\begin{subequations}
\begin{eqnarray}
b^{gg}({}^3\!P_J^{[8]})
=\frac{\pi ^3 \alpha_s^3}{m_c^5 s^2 t^2 \left(s-4 m_c^2\right)^4
\left(t-4 m_c^2\right)^4 (s+t)^4 \left(-4 m_c^2+s+t\right)^2}
\end{eqnarray}
\begin{eqnarray}
h_1&=&-s t \left(-4 m_c^2+s+t\right) \Big{[}1048576 m_c^{20}
\left(51 s^4+187 s^3 t+244 s^2 t^2+127 s t^3+51 t^4\right)\nonumber\\
&-&262144 m_c^{18} \left(306 s^5+1357 s^4 t+2375 s^3 t^2
+1905 s^2 t^3+1007 s t^4+306 t^5\right)\nonumber\\
&+&65536 m_c^{16} \left(911 s^6+4526 s^5 t+9668 s^4 t^2+10344 s^3 t^3
+7048 s^2 t^4+3356 s t^5+831 t^6\right)\nonumber\\
&-&32768 m_c^{14} (807 s^7+4441 s^6 t+10914 s^5 t^2
+14369 s^4 t^3+12229 s^3 t^4+7369 s^2 t^5\nonumber\\
&+&3126 s t^6+657 t^7)+8192 m_c^{12} (888 s^8+5519 s^7 t
+15179 s^6 t^2+23051 s^5 t^3+23276 s^4 t^4\nonumber\\
&+&17136 s^3 t^5+9319 s^2 t^6+3604 s t^7+648 t^8)-2048 m_c^{10}
(635 s^9+4541 s^8 t+13692 s^7 t^2\nonumber\\
&+&22822 s^6 t^3+25571 s^5 t^4+21816 s^4 t^5+14607 s^3 t^6
+7442 s^2 t^7+2701 s t^8+405 t^9)\nonumber\\
&+&256 m_c^8 (627 s^{10}+5195 s^9 t+16915 s^8 t^2
+29976 s^7 t^3+34868 s^6 t^4+31408 s^5 t^5\nonumber\\
&+&24488 s^4 t^6+16106 s^3 t^7+8015 s^2 t^8+2755 s t^9+327 t^{10})
-64 m_c^6 (210 s^{11}+2085 s^{10} t\nonumber\\
&+&7585 s^9 t^2+14784 s^8 t^3+18290 s^7 t^4+16790 s^6 t^5
+13650 s^5 t^6+10810 s^4 t^7+7164 s^3 t^8\nonumber\\
&+&3345 s^2 t^9+1005 s t^{10}+90 t^{11})+16 m_c^4 (35 s^{12}
+506 s^{11} t+2338 s^{10} t^2+5745 s^9 t^3\nonumber\\
&+&9126 s^8 t^4+10612 s^7 t^5+9988 s^6 t^6+8172 s^5 t^7+5746 s^4 t^8
+3105 s^3 t^9+1138 s^2 t^{10}\nonumber\\
&+&246 s t^{11}+15 t^{12})-4 m_c^2 s t \left(s^2+s t+t^2\right)^2
(46 s^7+270 s^6 t+611 s^5 t^2+761 s^4 t^3\nonumber\\
&+&681 s^3 t^4+451 s^2 t^5+170 s t^6+26 t^7)+11 s^2 t^2 (s+t)^2
\left(s^2+s t+t^2\right)^4\Big{]}
\end{eqnarray}
\begin{eqnarray}
h_2&=&32 m_c^2 \Big{[}4194304 s (s-t)^2 t m_c^{22}-262144 s t
\left(11 s^3-44 t s^2+11 t^2 s+6 t^3\right) m_c^{20}\nonumber\\
&-&65536 \left(76 s^6+237 t s^5+436 t^2 s^4-153 t^3 s^3-126 t^4 s^2
+118 t^5 s+36 t^6\right) m_c^{18}\nonumber\\
&+&16384 (416 s^7+1681 t s^6+2772 t^2 s^5+495 t^3 s^4
-892 t^4 s^3+346 t^5 s^2+998 t^6 s\nonumber\\
&+&288 t^7) m_c^{16}-4096 (940 s^8+4547 t s^7+8571 t^2 s^6
+4687 t^3 s^5-470 t^4 s^4+878 t^5 s^3\nonumber\\
&+&4017 t^6 s^2+3802 t^7 s+1008 t^8) m_c^{14}+1024 (1120 s^9
+6286 t s^8+13654 t^2 s^7+11090 t^3 s^6\nonumber\\
&+&3254 t^4 s^5+4751 t^5 s^4+10960 t^6 s^3+12707 t^7 s^2+8314 t^8 s
+2016 t^9) m_c^{12}-256 (740 s^{10}\nonumber\\
&+&4700 t s^9+11283 t^2 s^8+11038 t^3 s^7+5528 t^4 s^6
+9919 t^5 s^5+21427 t^6 s^4+26283 t^7 s^3\nonumber\\
&+&21360 t^8 s^2+11286 t^9 s+2520 t^{10}) m_c^{10}+64 (256 s^{11}
+1780 t s^{10}+4120 t^2 s^9+3090 t^3 s^8\nonumber\\
&+&544 t^4 s^7+7817 t^5 s^6+24064 t^6 s^5+34463 t^7 s^4+31912 t^8 s^3
+21376 t^9 s^2+9738 t^{10} s\nonumber\\
&+&2016 t^{11}) m_c^8-16 (36 s^{12}+244 t s^{11}-40 t^2 s^{10}
-2560 t^3 s^9-5744 t^4 s^8-1877 t^5 s^7\nonumber\\
&+&11279 t^6 s^6+24269 t^7 s^5+27618 t^8 s^4+21672 t^9 s^3+12769 t^{10} s^2
+5218 t^{11} s+1008 t^{12}) m_c^6\nonumber\\
&+&8 t (-6 s^{12}-192 t s^{11}-1024 t^2 s^{10}-2580 t^3 s^9
-3274 t^4 s^8-1283 t^5 s^7+2616 t^6 s^6\nonumber\\
&+&5395 t^7 s^5+5517 t^8 s^4+3840 t^9 s^3+2057 t^{10} s^2+786 t^{11} s
+144 t^{12}) m_c^4+t^2 (52 s^{12}\nonumber\\
&+&396 t s^{11}+1396 t^2 s^{10}+2700 t^3 s^9+2943 t^4 s^8+1412 t^5 s^7
-850 t^6 s^6-1920 t^7 s^5-1653 t^8 s^4\nonumber\\
&-&916 t^9 s^3-452 t^{10} s^2-188 t^{11} s-36 t^{12}) m_c^2
-s t^3 \left(s^2+t s+t^2\right)^2 (s^7+4 t s^6-4 t^2 s^5\nonumber\\
&-&26 t^3 s^4-22 t^4 s^3+4 t^5 s^2+9 t^6 s+2 t^7)\Big{]}
\end{eqnarray}
\begin{eqnarray}
h_3&=&32 m_c^2 \Big{[}262144 m_c^{18} s t \left(28 s^4+93 s^3 t
+126 s^2 t^2+73 s t^3+28 t^4\right)-16384 m_c^{16} s t (653 s^5\nonumber\\
&+&2802 s^4 t+4924 s^3 t^2+4254 s^2 t^3+2407 s t^4+688 t^5)
-4096 m_c^{14} (116 s^8-1263 s^7 t\nonumber\\
&-&8068 s^6 t^2-18556 s^5 t^3-21164 s^4 t^4-15711 s^3 t^5
-7768 s^2 t^6-1718 s t^7+36 t^8)\nonumber\\
&+&1024 m_c^{12} (656 s^9+369 s^8 t-9566 s^7 t^2
-32833 s^6 t^3-49800 s^5 t^4-46665 s^4 t^5\nonumber\\
&-&30158 s^3 t^6-12591 s^2 t^7-1916 s t^8+216 t^9)-256 m_c^{10}
(1540 s^{10}+5543 s^9 t+1997 s^8 t^2\nonumber\\
&-&22185 s^7 t^3-53859 s^6 t^4-66855 s^5 t^5-54509 s^4 t^6
-29995 s^3 t^7-9693 s^2 t^8-92 s t^9\nonumber\\
&+&540 t^{10})+64 m_c^8 (1920 s^{11}+9816 s^{10} t
+18914 s^9 t^2+13315 s^8 t^3-10042 s^7 t^4-33052 s^6 t^5\nonumber\\
&-&39132 s^5 t^6-27447 s^4 t^7-11120 s^3 t^8-576 s^2 t^9
+2236 s t^{10}+720 t^{11})-16 m_c^6 (1340 s^{12}\nonumber\\
&+&8240 s^{11} t+21729 s^{10} t^2+32953 s^9 t^3+32174 s^8 t^4
+21270 s^7 t^5+9410 s^6 t^6+3455 s^5 t^7\nonumber\\
&+&3594 s^4 t^8+4948 s^3 t^9+4629 s^2 t^{10}+2550 s t^{11}
+540 t^{12})+8 m_c^4 (248 s^{13}+1724s^{12} t\nonumber\\
&+&5352 s^{11} t^2+10237 s^{10} t^3+14018 s^9 t^4+15183 s^8 t^5
+13805 s^7 t^6+10710 s^6 t^7+7198 s^5 t^8\nonumber\\
&+&4463 s^4 t^9+2707 s^3 t^{10}+1507 s^2 t^{11}+604 s t^{12}
+108 t^{13})-m_c^2 (s+t)^2 (76 s^{12}+412 s^{11} t\nonumber\\
&+&1004 s^{10} t^2+1740 s^9 t^3+2541 s^8 t^4+3200 s^7 t^5+3202 s^6 t^6
+2320 s^5 t^7+1141 s^4 t^8\nonumber\\
&+&440 s^3 t^9+204 s^2 t^{10}+132 s t^{11}+36 t^{12})-s t (s+t)^3
\left(s^2+s t+t^2\right)^2 (2 s^6+5 s^5 t-8 s^4 t^2\nonumber\\
&-&23 s^3 t^3-8 s^2 t^4+5 s t^5+2 t^6)\Big{]}
\end{eqnarray}
\begin{eqnarray}
h_4&=&32 m_c^2 \Big{[}65536 s t \left(59 s^4+150 t s^3+252 t^2 s^2
+182 t^3 s+53 t^4\right) m_c^{18}-16384 (96 s^7\nonumber\\
&+&690 t s^6+1748 t^2 s^5+2626 t^3 s^4+2313 t^4 s^3+1154 t^5 s^2
+213 t^6 s-36 t^7) m_c^{16}\nonumber\\
&+&4096 (536 s^8+3159 t s^7+8087 t^2 s^6+12811 t^3 s^5
+12362 t^4 s^4+7490 t^5 s^3+2561 t^6 s^2\nonumber\\
&-&26 t^7 s-252 t^8) m_c^{14}-1024 (1240 s^9+7337 t s^8
+19822 t^2 s^7+33719 t^3 s^6+36841 t^4 s^5\nonumber\\
&+&25464 t^5 s^4+10399 t^6 s^3+1164 t^7 s^2-1706 t^8 s-756 t^9)
m_c^{12}+256 (1520 s^{10}+9504 t s^9\nonumber\\
&+&27735 t^2 s^8+50970 t^3 s^7+62054 t^4 s^6+49255 t^5 s^5
+22995 t^6 s^4+2530 t^7 s^3-4972 t^8 s^2\nonumber\\
&-&4387 t^9 s-1260 t^{10}) m_c^{10}-64 (1040 s^{11}
+6961 t s^{10}+22156 t^2 s^9+44286 t^3 s^8+58938 t^4 s^7\nonumber\\
&+&51617 t^5 s^6+26090 t^6 s^5+904 t^7 s^4-10831 t^8 s^3
-10450 t^9 s^2-5487 t^{10} s-1260 t^{11}) m_c^8\nonumber\\
&+&16 (376 s^{12}+2682 t s^{11}+9390 t^2 s^{10}+20757 t^3 s^9
+30044 t^4 s^8+27253 t^5 s^7+11565 t^6 s^6\nonumber\\
&-&5978 t^7 s^5-15058 t^8 s^4-14380 t^9 s^3-9099 t^{10} s^2
-3812 t^{11} s-756 t^{12}) m_c^6-4 (56 s^{13}\nonumber\\
&+&408 t s^{12}+1632 t^2 s^{11}+4317 t^3 s^{10}+7054 t^4 s^9
+5921 t^5 s^8-796 t^6 s^7-8944 t^7 s^6\nonumber\\
&-&12637 t^8 s^5-11090 t^9 s^4-7241 t^{10} s^3-3756 t^{11} s^2
-1396 t^{12} s-252 t^{13}) m_c^4-t (8 s^{13}\nonumber\\
&-&186 t^2 s^{11}-472 t^3 s^{10}-9 t^4 s^9+1935 t^5 s^8+4462 t^6 s^7
+5508 t^7 s^6+4381 t^8 s^5+2457 t^9 s^4\nonumber\\
&+&1116 t^{10} s^3+504 t^{11} s^2+196 t^{12} s+36 t^{13}) m_c^2-s t^2
\left(s^3+2 t s^2+2 t^2 s+t^3\right)^2\nonumber\\
&&\left(s^6+3 t s^5-9 t^2 s^4-22 t^3 s^3-6 t^4 s^2+6 t^5 s+2 t^6\right)\Big{]}
\end{eqnarray}
\end{subequations}

$g+g\to c\bar{c}({}^3\!S_1^{[8]},{}^3\!D_1^{[8]})+g$:
\begin{subequations}
\begin{eqnarray}
b^{gg}({}^3\!S_1^{[8]},{}^3\!D_1^{[8]})
=-\frac{\pi ^3 \alpha_s^3}{54 \sqrt{15} m_c^5 s t 
\left(s-4 m_c^2\right)^4\left(t-4 m_c^2\right)^4 (s+t)^4 \left(-4 m_c^2+s+t\right)}
\end{eqnarray}
\begin{eqnarray}
h_1&=&s t \left(s-4 m_c^2\right) \left(4 m_c^2-t\right) (s+t)
\left(4 m_c^2-s-t\right) \Big{[}49152 m_c^{16} (1261 s+1459 t)\nonumber\\
&-&1024 m_c^{14} \left(34157 s^2+40400 s t+42419 t^2\right)
+512 m_c^{12} (5978 s^3-29869 s^2 t-21472 s t^2\nonumber\\
&+&11999 t^3)+64 m_c^{10} \left(10841 s^4+214882 s^3 t+450086 s^2 t^2
+196198 s t^3+5225 t^4\right)\nonumber\\
&+&32 m_c^8 \left(19811 s^5-64363 s^4 t-264691 s^3 t^2-265735 s^2 t^3
-67162 s t^4+19244 t^5\right)\nonumber\\
&-&4 m_c^6 (86162 s^6+168328 s^5 t+43407 s^4 t^2
-172806 s^3 t^3+34263 s^2 t^4+163270 s t^5\nonumber\\
&+&85568 t^6)+m_c^4 (57024 s^7+220986 s^6 t
+406490 s^5 t^2+459356 s^4 t^3+459698 s^3 t^4\nonumber\\
&+&407102 s^2 t^5+221256 s t^6+57024 t^7)-162 m_c^2
(22 s^8+128 s^7 t+331 s^6 t^2+535 s^5 t^3\nonumber\\
&+&632 s^4 t^4+535 s^3 t^5+331 s^2 t^6+128 s t^7+22 t^8)
+405 s t (s+t) \left(s^2+s t+t^2\right)^3\Big{]}
\end{eqnarray}
\begin{eqnarray}
h_2&=&-2 m_c^2 t \left(4 m_c^2-t\right) \Big{[}131072 m_c^{18}
s (19027 s+38911 t)-32768 m_c^{16} (34301 s^3\nonumber\\
&+&175284 s^2 t+117007 s t^2+1620 t^3)-4096 m_c^{14}
(81504 s^4-265155 s^3 t-587452 s^2 t^2 \nonumber\\
&-&154537 s t^3-25920 t^4)+1024 m_c^{12} (340730 s^5
+674275 s^4 t+265516 s^3 t^2+143007 s^2 t^3\nonumber\\
&+&77548 s t^4-90720 t^5)-256 m_c^{10} (443414 s^6
+1706395 s^5 t+2270361 s^4 t^2+1548543 s^3 t^3\nonumber\\
&+&710607 s^2 t^4-161108 s t^5-181440 t^6)+64 m_c^8
(367548 s^7+1961043 s^6 t+3676724 s^5 t^2\nonumber\\
&+&3126628 s^4 t^3+1228754 s^3 t^4-268787 s^2 t^5-686126 s t^6
-223560 t^7)-16 m_c^6 (235756 s^8\nonumber\\
&+&1490664 s^7 t+3518089 s^6 t^2+3849434 s^5 t^3
+1701248 s^4 t^4-421485 s^3 t^5-1160629 s^2 t^6\nonumber\\
&-&731477 s t^7-168480 t^8)+4 m_c^4 (103680 s^9
+755956 s^8 t+2167956 s^7 t^2+3098173 s^6 t^3\nonumber\\
&+&2077386 s^5 t^4+205573 s^4 t^5-775916 s^3 t^6-761172 s^2 t^7
-345856 s t^8-71280 t^9)\nonumber\\
&-&m_c^2 (20736 s^{10}+207360 s^9 t+737224 s^8 t^2
+1361080 s^7 t^3+1370281 s^6 t^4+690475 s^5 t^5\nonumber\\
&+&23643 s^4 t^6-195151 s^3 t^7-151624 s^2 t^8-60912 s t^9
-12960 t^{10})+162 s t (32 s^9+160 s^8 t\nonumber\\
&+&359 s^7 t^2+476 s^6 t^3+409 s^5 t^4+233 s^4 t^5+96 s^3 t^6
+39 s^2 t^7+16 s t^8+4 t^9)\Big{]}
\end{eqnarray}
\begin{eqnarray}
h_3&=&-2 m_c^2 (s+t) \left(4 m_c^2-s-t\right) \Big{[}65536 m_c^{16} s t
(6701 s+6899 t)+8192 m_c^{14} (1620 s^4\nonumber\\
&-&63459 s^3 t-138920 s^2 t^2-70317 s t^3+1620 t^4)-1024 m_c^{12}
(22680 s^5-201895 s^4 t\nonumber\\
&-&495519 s^3 t^2-539529 s^2 t^3-244321 s t^4+22680 t^5)+256 m_c^{10}
(68040 s^6+28875 s^5 t\nonumber\\
&-&140260 s^4 t^2-209950 s^3 t^3-209740 s^2 t^4-717 s t^5+68040 t^6)
-64 m_c^8 (113400 s^7\nonumber\\
&+&353683 s^6 t+595357 s^5 t^2+986984 s^4 t^3+1022588 s^3 t^4
+637657 s^2 t^5+359587 s t^6\nonumber\\
&+&113400 t^7)+16 m_c^6 (110160 s^8+383177 s^7 t
+750828 s^6 t^2+1345133 s^5 t^3+1831816 s^4 t^4\nonumber\\
&+&1470557 s^3 t^5+821388 s^2 t^6+389549 s t^7+110160 t^8)
-4 m_c^4 (58320 s^9+205510 s^8 t\nonumber\\
&+&354804 s^7 t^2+556953 s^6 t^3+851841 s^5 t^4+878157 s^4 t^5
+598875 s^3 t^6+370788 s^2 t^7\nonumber\\
&+&205888 s t^8+58320 t^9)+m_c^2 (s+t)^2 (12960 s^8
+16848 s^7 t+262 s^6 t^2-23497 s^5 t^3\nonumber\\
&+&8468 s^4 t^4-26953 s^3 t^5-8 s^2 t^6+16848 s t^7+12960 t^8)+162 s t
(4 s^9+24 s^8 t+75 s^7 t^2\nonumber\\
&+&120 s^6 t^3+149 s^5 t^4+149 s^4 t^5+120 s^3 t^6
+75 s^2 t^7+24 s t^8+4 t^9)\Big{]}
\end{eqnarray}
\begin{eqnarray}
h_4&=&m_c^2 \Big{[}3145728 m_c^{20} s t (2357 s-3083 t)
-131072 m_c^{18} (1620 s^4+105283 s^3 t+8642 s^2 t^2\nonumber\\
&-&96641 s t^3-1620 t^4)+16384 m_c^{16} (25920 s^5
+509621 s^4 t+896167 s^3 t^2-85037 s^2 t^3\nonumber\\
&-&245935 s t^4-32400 t^5)-8192 m_c^{14} (45360 s^6
+320292 s^5 t+1056932 s^4 t^2+856843 s^3 t^3\nonumber\\
&+&146706 s^2 t^4+583 s t^5-71280 t^6)+1024 m_c^{12}
(181440 s^7+599428 s^6 t+1988277 s^5 t^2\nonumber\\
&+&2840253 s^4 t^3+1083297 s^3 t^4+141039 s^2 t^5
-376166 s t^6-362880 t^7)\nonumber\\
&-&256 m_c^{10} (223560 s^8+623352 s^7 t
+977271 s^6 t^2+992272 s^5 t^3-1796934 s^4 t^4\nonumber\\
&-&3815074 s^3 t^5-2849597 s^2 t^6-1832690 s t^7-586440 t^8)
+64 m_c^8 (168480 s^9+592859 s^8 t\nonumber\\
&+&679396 s^7 t^2-438617 s^6 t^3-4960050 s^5 t^4-10323584 s^4 t^5
-9863307 s^3 t^6-6026378 s^2 t^7\nonumber\\
&-&2744639 s t^8-615600 t^9)-16 m_c^6 (71280 s^{10}
+312714 s^9 t+668005 s^8 t^2+413842 s^7 t^3\nonumber\\
&-&2602893 s^6 t^4-8848230 s^5 t^5-12123493 s^4 t^6-9435072 s^3 t^7
-5216495 s^2 t^8-2055090 s t^9\nonumber\\
&-&408240 t^{10})+4 m_c^4 (12960 s^{11}+49248 s^{10} t
+294940 s^9 t^2+872129 s^8 t^3+834712 s^7 t^4\nonumber\\
&-&1612967 s^6 t^5-5014855 s^5 t^6-5770958 s^4 t^7-4063097 s^3 t^8
-2154136 s^2 t^9-805760 s t^{10}\nonumber\\
&-&155520 t^{11})+m_c^2 t (7776 s^{11}-55728 s^{10} t
-413458 s^9 t^2-1003935 s^8 t^3-1092466 s^7 t^4\nonumber\\
&-&310362 s^6 t^5+540758 s^5 t^6+744113 s^4 t^7+542710 s^3 t^8
+309728 s^2 t^9+124416 s t^{10}\nonumber\\
&+&25920 t^{11})+324 s t^2 (s+t)^2 (8 s^8+44 s^7 t+83 s^6 t^2
+85 s^5 t^3+64 s^4 t^4+37 s^3 t^5+31 s^2 t^6\nonumber\\
&+&16 s t^7+4 t^8)\Big{]}
\end{eqnarray}
\end{subequations}

$q(\bar{q})+g\to c\bar{c}({}^3\!P_1^{[1]})+q(\bar{q})$:
\begin{subequations}
\begin{eqnarray}
a^{q(\bar{q})g}({}^3\!P_1^{[1]})=\frac{32 \pi ^3 \alpha_s^3}{81 m_c^3 (s+t)^4}
\end{eqnarray}
\begin{eqnarray}
g_1=-(s+t) \left[-4 m_c^2 (s+t)+s^2+t^2\right]
\end{eqnarray}
\begin{eqnarray}
g_2=-16 m_c^2 s
\end{eqnarray}
\begin{eqnarray}
g_3=0
\end{eqnarray}
\begin{eqnarray}
g_4=-8 m_c^2 (s-t)
\end{eqnarray}
\begin{eqnarray}
b^{q(\bar{q})g}({}^3\!P_1^{[1]})
=\frac{16 \pi ^3 \alpha_s^3}{405 m_c^5 \left(4 m_c^2-s\right) (s+t)^5}
\end{eqnarray}
\begin{eqnarray}
h_1&=&-(s+t) \Big{[}16 m_c^4 \left(s^2+22 s t+21 t^2\right)-4 m_c^2 \left(12 s^3
+43 s^2 t-8 s t^2+21 t^3\right)\nonumber\\
&+&11 s \left(s^3+s^2 t+s t^2+t^3\right)\Big{]}
\end{eqnarray}
\begin{eqnarray}
h_2=16 m_c^2 s \left[4 m_c^2 (s+41 t)-s (s+t)\right]
\end{eqnarray}
\begin{eqnarray}
h_3=0
\end{eqnarray}
\begin{eqnarray}
h_4=8 m_c^2 \left[4 m_c^2 \left(s^2+50 s t-31 t^2\right)-s^3+s t^2\right]
\end{eqnarray}
\end{subequations}

$q(\bar{q})+g\to c\bar{c}({}^3\!P_2^{[1]})+q(\bar{q})$:
\begin{subequations}
\begin{eqnarray}
a^{q(\bar{q})g}({}^3\!P_2^{[1]})=-\frac{32 \pi ^3 \alpha_s^3}{1215 m_c^3
(s+t)^4 \left(-4 m_c^2+s+t\right)}
\end{eqnarray}
\begin{eqnarray}
g_1=16 m_c^4 \left(19 s^2+30 s t+19 t^2\right)-8 m_c^2
\left(s^3+16 s^2 t+16 s t^2+t^3\right)+(s+t)^2 \left(s^2+t^2\right)\nonumber\\
\end{eqnarray}
\begin{eqnarray}
g_2=48 m_c^2 \left(32 m_c^4+4 m_c^2 (s+2 t)-s^2-s t+2 t^2\right)
\end{eqnarray}
\begin{eqnarray}
g_3=96 m_c^2 (s+t)^2
\end{eqnarray}
\begin{eqnarray}
g_4=-24 m_c^2 \left(4 m_c^2 (s-t)+s^2-4 s t-5 t^2\right)
\end{eqnarray}
\begin{eqnarray}
b^{q(\bar{q})g}({}^3\!P_2^{[1]})
=\frac{16 \pi ^3 \alpha_s^3}{6075 m_c^5 \left(4 m_c^2-s\right)
(s+t)^5 \left(-4 m_c^2+s+t\right)^2}
\end{eqnarray}
\begin{eqnarray}
h_1&=&-\left(4 m_c^2-s-t\right) \Big{[}320 m_c^6 \left(19 s^3
+141 s^2 t+153 s t^2+95 t^3\right)\nonumber\\
&+&16 m_c^4 (73 s^4+39 s^3 t-923 s^2 t^2-931 s t^3-42 t^4)\nonumber\\
&+&4 m_c^2 (s+t)^2 \left(19 s^3+139 s^2 t-s t^2+17 t^3\right)
-7 s (s+t)^3 \left(s^2+t^2\right)\Big{]}
\end{eqnarray}
\begin{eqnarray}
h_2&=&48 m_c^2 \Big{[}2560 m_c^8 (s-7 t)-64 m_c^6
\left(49 s^2-3 s t+4 t^2\right)\nonumber\\
&+&16 m_c^4 \left(37 s^3+113 s^2 t+180 s t^2+8 t^3\right)+4 m_c^2
\left(5 s^4-21 s^3 t-69 s^2 t^2+23 s t^3+66 t^4\right)\nonumber\\
&-&s (s+t)^2 \left(3 s^2-s t+6 t^2\right)\Big{]}
\end{eqnarray}
\begin{eqnarray}
h_3&=&-96 m_c^2 (s+t)^2 \Big{[}80 m_c^4 (3 s+7 t)-4 m_c^2
\left(16 s^2+47 s t+31 t^2\right)+s (s+t)^2\Big{]}
\end{eqnarray}
\begin{eqnarray}
h_4&=&24 m_c^2 \Big{[}320 m_c^6 \left(s^2+10 s t-7 t^2\right)
-80 m_c^4 \left(s^3+8 s^2 t+31 s t^2+24 t^3\right)\nonumber\\
&-&4 m_c^2 (s+t)^2 \left(s^2+26 s t-155 t^2\right)+s (s-5 t) (s+t)^3\Big{]}
\end{eqnarray}
\end{subequations}

$q(\bar{q})+g\to c\bar{c}({}^3\!S_1^{[8]})+q(\bar{q})$:
\begin{subequations}
\begin{eqnarray}
b^{q(\bar{q})g}({}^3\!S_1^{[8]})=\frac{\pi ^3 \alpha_s^3}{162 m_c^5 s t 
\left(s-4 m_c^2\right)(s+t)^3 \left(-4 m_c^2+s+t\right)}
\end{eqnarray}
\begin{eqnarray}
h_1&=&-\left(4 m_c^2-s-t\right) \Big{[}128 m_c^6 \left(20 s^3+69 s^2 t
-39 s t^2+20 t^3\right)-32 m_c^4 (40 s^4+104 s^3 t\nonumber\\
&+&45 s^2 t^2+s t^3+20 t^4)+4 m_c^2 \left(108 s^5+175 s^4 t+77 s^3 t^2
+207 s^2 t^3+s t^4+20 t^5\right)\nonumber\\
&-&11 s \left(4 s^5+3 s^4 t+7 s^3 t^2+7 s^2 t^3+3 s t^4+4 t^5\right)\Big{]}
\end{eqnarray}
\begin{eqnarray}
h_2&=&-16 m_c^2 \Big{[}576 m_c^6 (s-t)^2+16 m_c^4 \left(26 s^3
+105 s^2 t-3 s t^2+26 t^3\right)-4 m_c^2 (55 s^4\nonumber\\
&+&155 s^3 t+99 s^2 t^2+34 s t^3+35 t^4)+s \left(20 s^4+44 s^3 t+21 s^2 t^2
+44 s t^3+11 t^4\right)\Big{]}
\end{eqnarray}
\begin{eqnarray}
h_3&=&-32 m_c^2 \left(4 m_c^2-s-t\right) \Big{[}4 m_c^2
(44 s^3+87 s^2 t-21 s t^2+44 t^3)\nonumber\\
&-&5 s \left(4 s^3+3 s^2 t+3 s t^2+4 t^3\right)\Big{]}
\end{eqnarray}
\begin{eqnarray}
h_4&=&-8 m_c^2 \Big{[}16 m_c^4 \left(97 s^3+147 s^2 t-15 s t^2
+79 t^3\right)-4 m_c^2 (146 s^4+265 s^3 t+189 s^2 t^2\nonumber\\
&+&77 s t^3+79 t^4)+s \left(49 s^4+70 s^3 t+60 s^2 t^2+70 s t^3+31 t^4\right)\Big{]}
\end{eqnarray}
\end{subequations}

$q(\bar{q})+g\to c\bar{c}({}^3\!P_J^{[8]})+q(\bar{q})$:
\begin{eqnarray}
b^{q(\bar{q})g}({}^3\!P_J^{[8]})
=\frac{15\alpha_s}{128\alpha}b^{q(\bar{q})\gamma}({}^3\!P_J^{[8]})
\end{eqnarray}
The coefficients $h_k$ are the same as those in Eq.~(\ref{eq:fifteen}).

$q(\bar{q})+g\to c\bar{c}({}^3\!S_1^{[8]},{}^3\!D_1^{[8]})+q(\bar{q})$:
\begin{subequations}
\begin{eqnarray}
b^{q(\bar{q})g}({}^3\!S_1^{[8]},{}^3\!D_1^{[8]})
=-\frac{\pi ^3 \alpha_s^3}{27 \sqrt{15} m_c^5 s t (s+t)^4
\left(-4 m_c^2+s+t\right)}
\end{eqnarray}
\begin{eqnarray}
h_1&=&-(s+t) \left(-4 m_c^2+s+t\right) \Big{[}160 m_c^4 \left(4 s^3
+3 s^2 t+3 s t^2+4 t^3\right)-4 m_c^2 (40 s^4+97 s^3 t\nonumber\\
&+&6 s^2 t^2+97 s t^3+40 t^4)+5 \left(4 s^5+3 s^4 t+7 s^3 t^2+7 s^2 t^3+3 s t^4+4 t^5\right)\Big{]}
\end{eqnarray}
\begin{eqnarray}
h_2&=&8 m_c^2 \Big{[}3456 m_c^6 (s-t)^2-144 m_c^4 \left(4 s^3
-s^2 t-16 s t^2+13 t^3\right)+4 m_c^2 (22 s^4+25 s^3 t\nonumber\\
&+&114 s^2 t^2+7 s t^3+112 t^4)-40 s^5-128 s^4 t-157 s^3 t^2-103 s^2 t^3-83 s t^4-49 t^5\Big{]}
\end{eqnarray}
\begin{eqnarray}
h_3&=&8 m_c^2 (s+t)^2 \left(107 s^2-74 s t+107 t^2\right) \left(4 m_c^2-s-t\right)
\end{eqnarray}
\begin{eqnarray}
h_4&=&8 m_c^2 \Big{[}216 m_c^4 \left(3 s^3-5 s^2 t+5 s t^2-3 t^3\right)
+2 m_c^2 (17 s^4+158 s^3 t+66 s^2 t^2\nonumber\\
&+&122 s t^3+197 t^4)-(s+t)^2 \left(49 s^3+48 s^2 t-15 s t^2+58 t^3\right)\Big{]}
\end{eqnarray}
\end{subequations}

$\bar{q}+q\to c\bar{c}({}^3\!P_1^{[1]})+g$:
\begin{subequations}
\begin{eqnarray}
a^{\bar{q}q}({}^3\!P_1^{[1]})
=\frac{256 \pi ^3 \alpha_s^3}{243 m_c^3 \left(s-4 m_c^2\right)^4}
\end{eqnarray}
\begin{eqnarray}
g_1=-\left(4 m_c^2-s\right) \left(4 m_c^2 (s+2 t)-s^2-2 s t-2 t^2\right)
\end{eqnarray}
\begin{eqnarray}
g_2=16 m_c^2 \left(4 m_c^2-s-t\right)
\end{eqnarray}
\begin{eqnarray}
g_3=16 m_c^2 t
\end{eqnarray}
\begin{eqnarray}
g_4=32 m_c^4-8 m_c^2 s
\end{eqnarray}
\begin{eqnarray}
b^{\bar{q}q}({}^3\!P_1^{[1]})
=\frac{128 \pi ^3 \alpha_s^3}{1215 m_c^5 \left(4 m_c^2-s\right)^5}
\end{eqnarray}
\begin{eqnarray}
h_1=\left(4 m_c^2-s\right) \big{[}16 m_c^4 (31 s+42 t)-8 m_c^2
\left(21 s^2+32 s t+21 t^2\right)+11 s \left(s^2+2 s t+2 t^2\right)\big{]}\nonumber\\
\end{eqnarray}
\begin{eqnarray}
h_2=-16 m_c^2 \left(124 m_c^2-s\right) \left(4 m_c^2-s-t\right)
\end{eqnarray}
\begin{eqnarray}
h_3=16 m_c^2 t \left(s-124 m_c^2\right)
\end{eqnarray}
\begin{eqnarray}
h_4=-8 m_c^2 \left(4 m_c^2-s\right) \left(124 m_c^2-s\right)
\end{eqnarray}
\end{subequations}

$\bar{q}+q\to c\bar{c}({}^3\!P_2^{[1]})+g$:
\begin{subequations}
\begin{eqnarray}
a^{\bar{q}q}({}^3\!P_2^{[1]})
=\frac{256 \pi ^3 \alpha_s^3}{3645 m_c^3 s \left(s-4 m_c^2\right)^4}
\end{eqnarray}
\begin{eqnarray}
g_1&=&-4608 m_c^8+2304 m_c^6 (s+t)-16 m_c^4
\left(19 s^2+66 s t+36 t^2\right)
\nonumber\\
&+&8 m_c^2 s \left(s^2+16 s t+15 t^2\right)-s^2
\left(s^2+2 s t+2 t^2\right)
\end{eqnarray}
\begin{eqnarray}
g_2=-48 m_c^2 \left(32 m_c^4+4 m_c^2 (s+2 t)-s^2-s t+2 t^2\right)
\end{eqnarray}
\begin{eqnarray}
g_3=-48 m_c^2 \left(96 m_c^4-24 m_c^2 (s+t)+2 s^2+5 s t+2 t^2\right)
\end{eqnarray}
\begin{eqnarray}
g_4=-24 m_c^2 \left(96 m_c^4-4 m_c^2 (s+4 t)-s^2+4 s t+4 t^2\right)
\end{eqnarray}
\begin{eqnarray}
b^{\bar{q}q}({}^3\!P_2^{[1]})
=\frac{128 \pi ^3 \alpha_s^3}{18225 m_c^5 s^2 \left(4 m_c^2-s\right)^5}
\end{eqnarray}
\begin{eqnarray}
h_1&=&s \Big{[}276480 m_c^{10}-512 m_c^8 (221 s+246 t)
+64 m_c^6 \left(93 s^2+950 s t+492 t^2\right)\nonumber\\
&+&16 m_c^4 s \left(53 s^2-598 s t-458 t^2\right)+4 m_c^2 s^2
\left(33 s^2+154 s t+140 t^2\right)\nonumber\\
&-&7 s^3 \left(s^2+2 s t+2 t^2\right)\Big{]}
\end{eqnarray}
\begin{eqnarray}
h_2&=&48 m_c^2 \Big{[}2048 m_c^8+128 m_c^6 (19 s-8 t)
+16 m_c^4 \left(57 s^2+54 s t+8 t^2\right)\nonumber\\
&-&4 m_c^2 s \left(20 s^2+s t-58 t^2\right)
-s^2 \left(3 s^2+7 s t+10 t^2\right)\Big{]}
\end{eqnarray}
\begin{eqnarray}
h_3&=&48 m_c^2 \Big{[}9600 m_c^6 s-32 m_c^4
\left(58 s^2+77 s t-4 t^2\right)\nonumber\\
&+&4 m_c^2 s \left(52 s^2+137 s t+58 t^2\right)
-s^2 \left(6 s^2+13 s t+10 t^2\right)\Big{]}
\end{eqnarray}
\begin{eqnarray}
h_4&=&24 m_c^2 \Big{[}128 m_c^6 (85 s-8 t)+16 m_c^4
\left(3 s^2-100 s t+16 t^2\right)\nonumber\\
&-&8 m_c^2 s \left(3 s^2-68 s t-58 t^2\right)
-s^2 \left(7 s^2+20 s t+20 t^2\right)\Big{]}
\end{eqnarray}
\end{subequations}

$\bar{q}+q \to c\bar{c}({}^3\!S_1^{[8]})+g$:
\begin{subequations}
\begin{eqnarray}
b^{\bar{q}q}({}^3\!S_1^{[8]})
=\frac{4 \pi ^3 \alpha_s^3}{243 s t m_c^5 \left( s-4 m_c^2\right)^3
\left(-4 m_c^2+s+t\right)}
\end{eqnarray}
\begin{eqnarray}
h_1&=&s \big{[}64 m_c^4-4 m_c^2 (8 s+9 t)+4 s^2+9 s t+9 t^2\big{]}
\Big{[}704 m_c^6+16 m_c^4 (s-22 t)\nonumber\\
&+&4 m_c^2 \left(23 s^2+44 s t+22 t^2\right)-11 s \left(s^2+2 s t+2 t^2\right)\Big{]}
\end{eqnarray}
\begin{eqnarray}
h_2&=&16 m_c^2 \Big{[}64 m_c^6 (77 s-18 t)-96 m_c^4
\left(29 s^2+24 s t-12 t^2\right)\nonumber\\
&+&36 m_c^2 \left(13 s^3+22 s^2 t+9 s t^2-10 t^3\right)-20 s^4
-36 s^3 t-9 s^2 t^2+54 s t^3+36 t^4\Big{]}
\end{eqnarray}
\begin{eqnarray}
h_3&=&16 m_c^2 \Big{[}4352 m_c^6 s-48 m_c^4
\left(49 s^2+48 s t-6 t^2\right)\nonumber\\
&+&36 m_c^2 \left(10 s^3+16 s^2 t+9 s t^2-6 t^3\right)
-11 s^4+45 s^2 t^2+90 s t^3+36 t^4\Big{]}
\end{eqnarray}
\begin{eqnarray}
h_4&=&72 m_c^2 \left(-4 m_c^2+s+2 t\right)^2
\left(4 m_c^2 (s-2 t)+s^2+2 s t+2 t^2\right)
\end{eqnarray}
\end{subequations}

$\bar{q}+q\to c\bar{c}({}^3\!P_J^{[8]})+g$:
\begin{subequations}
\begin{eqnarray}
b^{\bar{q}q}({}^3\!P_J^{[8]})
=\frac{16 \pi ^3 \alpha_s^3}{27 m_c^5 s^2 \left(s-4 m_c^2\right)^4}
\end{eqnarray}
\begin{eqnarray}
h_1&=&s \left(4 m_c^2-s\right) \Big{[}3264 m_c^6
+16 m_c^4 (7 s-30 t)\nonumber\\
&+&4 m_c^2 \left(33 s^2+52 s t+30 t^2\right)
-11 s \left(s^2+2 s t+2 t^2\right)\Big{]}
\end{eqnarray}
\begin{eqnarray}
h_2&=&32 m_c^2 \Big{[}1152 m_c^6+48 m_c^4 (5 s-12 t)
+8 m_c^2 t (5 s+9 t)+s \left(s^2+2 s t+4 t^2\right)\Big{]}
\end{eqnarray}
\begin{eqnarray}
h_3&=&32 m_c^2 \Big{[}464 m_c^4 s+8 m_c^2
\left(s^2+9 s t+9 t^2\right)+s \left(3 s^2+6 s t+4 t^2\right)\Big{]}
\end{eqnarray}
\begin{eqnarray}
h_4&=&32 m_c^2 \Big{[}16 m_c^4 (19 s-18 t)+8 m_c^2
\left(7 s^2+7 s t+9 t^2\right)+s (s+2 t)^2\Big{]}
\end{eqnarray}
\end{subequations}

$\bar{q}+q \to c\bar{c}({}^3\!S_1^{[8]},{}^3\!D_1^{[8]})+g$:
\begin{subequations}
\begin{eqnarray}
b^{\bar{q}q}({}^3\!S_1^{[8]},{}^3\!D_1^{[8]})
=-\frac{8 \pi ^3 \alpha_s^3}{81 \sqrt{15} m_c^5 s t
\left(s-4 m_c^2\right)^4 \left(-4 m_c^2+s+t\right)}
\end{eqnarray}
\begin{eqnarray}
h_1&=&s \left(s-4 m_c^2\right) \Big{[}20480 m_c^{10}
-1024 m_c^8 (15 s+28 t)+64 m_c^6 (80 s^2+331 s t\nonumber\\
&+&310 t^2)-16 m_c^4 \left(80 s^3+411 s^2 t+705 s t^2+396 t^3\right)
+4 m_c^2 (60 s^4+277 s^3 t+570 s^2 t^2\nonumber\\
&+&576 s t^3+198 t^4)-5 s \left(4 s^4+17 s^3 t+35 s^2 t^2+36 s t^3+18 t^4\right)\Big{]}
\end{eqnarray}
\begin{eqnarray}
h_2&=&-8 m_c^2 \Big{[}512 m_c^8 (56 s-45 t)-64 m_c^6
\left(322 s^2+117 s t-360 t^2\right)+48 m_c^4 (116 s^3\nonumber\\
&+&120 s^2 t-75 s t^2-150 t^3)+m_c^2 \left(-712 s^4-900 s^3 t
+144 s^2 t^2+936 s t^3+720 t^4\right)\nonumber\\
&+&40 s^5+72 s^4 t+45 s^3 t^2+36 s t^4\Big{]}
\end{eqnarray}
\begin{eqnarray}
h_3&=&-8 m_c^2 \Big{[}10240 m_c^8 s-64 m_c^6 (115 s^2+99 s t
-90 t^2)+48 m_c^4 (53 s^3+48 s^2 t-39 s t^2\nonumber\\
&-&90 t^3)-4 m_c^2 \left(133 s^4+207 s^3 t+18 s^2 t^2
-342 s t^3-180 t^4\right)\nonumber\\
&+&s \left(49 s^4+162 s^3 t+261 s^2 t^2+144 s t^3+36 t^4\right)\Big{]}
\end{eqnarray}
\begin{eqnarray}
h_4&=&-72 m_c^2 \left(-4 m_c^2+s+2 t\right)^2 \Big{[}40 m_c^4
(s-2 t)+2 m_c^2 \left(9 s^2+8 s t+10 t^2\right)\nonumber\\
&+&s \left(-s^2+s t+t^2\right)\Big{]}
\end{eqnarray}
\end{subequations}

\end{appendix}

\end{document}